\documentclass[aps,prd,twocolumn,floatfix,nofootinbib,showpacs,superscriptaddress,tightenlines]{revtex4}

\usepackage{amsmath}
\usepackage{dcolumn}
\usepackage{lipsum}
\usepackage{amssymb}
\usepackage{url}
\usepackage{epsfig}
\usepackage{graphicx}
\usepackage{amsmath}
\usepackage{bm}
\usepackage{setspace}
\usepackage{lscape}
\usepackage{amsthm}
\usepackage{bbold}
\usepackage{dcolumn}
\usepackage{epsfig}
\usepackage{graphics}
\usepackage{graphicx}
\usepackage{longtable}
\usepackage{color}
\usepackage{xcolor}
\usepackage{bm}
\usepackage{xspace}
\usepackage{cancel}
\usepackage{float}
\usepackage{multirow}
\usepackage[mathscr,scaled=1.15]{urwchancal}
\DeclareFontFamily{OT1}{pzc}{}
\DeclareFontShape{OT1}{pzc}{m}{it}%
{<-> s * [1.15] pzcmi7t}{}
\DeclareMathAlphabet{\mathpzc}{OT1}{pzc}{m}{it}

\definecolor{darkgreen}{rgb}{0,0.5,0}
\definecolor{amber}{rgb}{1.0, 0.75, 0.0}
\definecolor{purple}{rgb}{0.5,0,0.5}
\definecolor{nblue}{rgb}{0.0,0.0,0.50}
\definecolor{scarlet}{rgb}{1.0,0.2,0}
\definecolor{darkmagenta}{rgb}{0.55, 0.0, 0.55}
\definecolor{darkolivegreen}{rgb}{0.33, 0.42, 0.18}
\definecolor{darkcandyapplered}{rgb}{0.64, 0.0, 0.0}
\definecolor{warmblack}{rgb}{0.0, 0.26, 0.26}
\definecolor{oxfordblue}{rgb}{0.0, 0.13, 0.28}
\definecolor{cyan(process)}{rgb}{0.0, 0.55, 0.55}
\definecolor{almond}{rgb}{0.94, 0.87, 0.8}
\definecolor{antiquewhite}{rgb}{0.98, 0.92, 0.84}
\definecolor{eggshell}{rgb}{0.94, 0.92, 0.84}
\definecolor{floralwhite}{rgb}{1.0, 0.98, 0.94}
\definecolor{linen}{rgb}{0.98, 0.94, 0.9}

\usepackage[colorlinks=true, pdfstartview=FitV, linkcolor=purple, citecolor= purple, urlcolor=blue]{hyperref}

\newcommand{\be}{\begin{equation}}
\newcommand{\tu}{\textcolor{red}{u}}

\newcommand{\fu}{\textcolor{blue}{\bar{f_2}}}
\newcommand{\fd}{\textcolor{blue}{f_1}}
\newcommand{\fdu}{\textcolor{blue}{f_2}}

\newcommand{\Me}{\textcolor{blue}{V}}
\newcommand{\Meps}{\textcolor{blue}{{PS}}}
\newcommand{\Dps}{\textcolor{blue}{{DPS}}}
\newcommand{\Mv}{\textcolor{blue}{V}}
\newcommand{\Mav}{\textcolor{blue}{AV}}
\newcommand{\Ms}{\textcolor{blue}{S}}
\newcommand{\Ds}{\textcolor{blue}{{DS}}}
\newcommand{\Dv}{\textcolor{blue}{{DV}}}
\newcommand{\Dav}{\textcolor{blue}{{DAV}}}

\newcommand{\Jpsi}{\textcolor{blue}{J/\Psi}}

\newcommand{\g}{\textcolor{blue}{\Gamma}}
\newcommand{\D}{\textcolor{magenta}{\Delta}}
\newcommand{\td}{\textcolor{darkcandyapplered}{d}}
\newcommand{\tb}{\textcolor{blue}{b}}
\newcommand{\tc}{\textcolor{darkmagenta}{c}}
\newcommand{\ts}{\textcolor{darkgreen}{s}}
\newcommand{\ee}{\end{equation}}
\newcommand{\bea}{\begin{eqnarray}}
\newcommand{\eea}{\end{eqnarray}}
\newcommand{\beas}{\begin{eqnarray*}}
\newcommand{\eeas}{\end{eqnarray*}}
\newcommand{\nn}{\nonumber}

\newcommand{\tq}{\textcolor{red}{q}}
\newcommand{\tqu}{\textcolor{blue}{q_1}}

\newcommand{\MeV}{\text{MeV}} 
\newcommand{\GeV}{\text{GeV}} 
\newcommand{\rmh}{\hat{\alpha}_{\mathrm {IR}}}

\newcommand{\gc}{\textcolor{darkgreen}{\bf{\gamma_5}}}
\newcommand{\mn}{\textcolor{darkgreen}{-}}
\newcommand{\mni}{\textcolor{darkgreen}{_{1^-}}}
\newcommand{\mpo}{\textcolor{darkgreen}{_{1^+}}}
\newcommand{\noo}{\textcolor{darkgreen}{_{0^-}}}
\newcommand{\nop}{\textcolor{darkgreen}{_{0^+}}}

\begin{document}
\title{Mesons and Baryons: Parity Partners}

\author{L.X. Guti\'errez-Guerrero}
\email[]{lxgutierrez@mctp.mx}
\thanks{}
\affiliation{CONACyT-Mesoamerican Centre for Theoretical Physics,
Universidad Aut\'onoma de Chiapas, Carretera Zapata Km. 4, Real
del Bosque (Ter\'an), Tuxtla Guti\'errez 29040, Chiapas, M\'exico}

\author{G. Paredes-Torres}
\email[]{gustavo.paredes@umich.mx}
\thanks{}
\affiliation{Instituto de F\'isica y Matem\'aticas, Universidad
Michoacana de San Nicol\'as de Hidalgo, Edificio C-3, Ciudad
Universitaria, Morelia, Michoac\'an 58040, M\'exico}

\author{A. Bashir}
\email[]{adnan.bashir@umich.mx}
\thanks{}
\affiliation{Instituto de F\'isica y Matem\'aticas, Universidad
Michoacana de San Nicol\'as de Hidalgo, Edificio C-3, Ciudad
Universitaria, Morelia, Michoac\'an 58040, M\'exico}


\begin{abstract}
We calculate masses of light and heavy mesons as well as baryons of negative parity containing $\tu,\td,\ts,\tc$ and $\tb$ quarks. It is an extension of our previous work where we had studied the  positive parity
baryons. We adopt a quark-diquark picture of baryons where the diquarks are non-pointlike
with a finite spatial extension.
The mathematical foundation for this analysis is implemented through a
symmetry-preserving Schwinger-Dyson equations treatment of a vector-vector contact interaction, which
preserves key features of quantum chromodynamics, such as confinement, chiral symmetry breaking, axial vector
Ward-Takahashi identity and low-energy Goldberger-Treiman relations.
This treatment  simultaneously describes mesons and provides attractive correlations for diquarks in the
$\overline{3}$ representation.
Employing this model, we  compute the spectrum and masses of all spin-1/2 and spin-3/2 baryons of negative parity, supplementing our earlier evaluation of positive parity baryons, containing 1, 2 or 3 heavy quarks. In the process, we calculate masses of a multitude of mesons and corresponding diquarks. Wherever possible, we make comparisons of our results with known experimental observations as well as theoretical predictions of several models and approaches including lattice quantum chromodynamics, finding satisfactory agreement. We also make predictions for heavier states not yet observed in the experiment.

\end{abstract}


\maketitle

\section{Introduction}

The heavy baryons are an immediate prediction of the quark model and their spectroscopy has
attracted a lot of attention in recent years due to their ongoing and expected observations in particle colliders
such as LHCb and Belle II. However, as has so often been the case for quantum chromodynamics (QCD) and hadron physics, what is relatively easier to calculate
theoretically is harder to measure experimentally and vice versa. For example, QCD based computation of the properties of
triply heavy baryons does not involve the complexities of the light quark dynamics but  there exists no experimental signal for any of them. Any realistic estimate of their production
shows that it is a wild goose chase in foreseeable future.

A baryon containing two charm quarks  $\Xi_{cc}^{++}$(3621.2 $\pm$ 0.7 MeV) was detected in the LHCb experiment at the Large
Hadron Collider (LHC) at CERN in proton collisions in both the 7TeV and 13TeV runs,~\cite{Aaij:2018gfl}.
It has made its entry into the Particle Data Group~\cite{Zyla:2020zbs}. A controversial double charm baryon $\Xi_{cc}^{+}$(3519 $\pm$ 2 MeV) was reported earlier by the SELEX Collaboration~\cite{Mattson:2002vu}.
This state remains unconfirmed by FOCUS~\cite{Ratti:2003ez}, BABAR~\cite{Aubert:2006qw}, Belle~\cite{Chistov:2006zj}
and LHCb experiments~\cite{Chatrchyan:2013rla} which did not find evidence for a state with the properties reported by SELEX.
However, these null results do not rule out the original observations,~\cite{Brodsky:2017ntu}. There are no
doubly heavy baryons observed with two bottom quarks or with one charm and one bottom quark. Several singly heavy quark baryons have been found, see~\cite{Zyla:2020zbs} for a detailed list. In this article, we set out to calculate the masses of negative parity baryons of spin 1/2 and 3/2. These are parity partners of the baryons we studied in Ref.~\cite{Gutierrez-Guerrero:2019uwa}. We employ a coupled analysis of Schwinger-Dyson equations (SDE), Bethe-Salpeter equations (BSE) describing the two-body bound-state problem, and Faddeev equations (FE) for a bound-state of three particles. The computations have been carried out within a vector-vector contact interaction (CI) which preserves essential features of QCD such as confinement, dynamical chiral symmetry breaking (DCSB) and the low energy implications of the divergence of the weak axial vector current.

 In relativistic quantum field theory, the parity partner of any given state can be obtained via a simple chiral rotation of the original state. If it were a good symmetry, parity partners would be of equal masses.
 However, the spectrum of mesons and baryons explicitly violates this symmetry.
 For example, in the light mesons sector, mass of $\rho_{1^-}$(770MeV) and that of its parity partner ${a_1}_{1^+}$(1260MeV)
 are not the same. The mass difference is of the order of 500MeV. In the baryon sector, similar amount of mass splitting is observed between
 the parity partners $\Delta_{3/2^+}$(1232MeV) and $\Delta_{3/2^-}$(1700MeV), as well as the nucleon $N_{1/2^+}$(939MeV) and $N^{*}_{1/2^-}(1535)$. This splitting is due to the effects of the DCSB-dictated repulsion involving
 P-wave components of the bound-state wave-functions, incorporated in the kernels of BSE and FE for the negative-parity hadrons.

%
The chiral-partners structure of hadrons including one heavy quark has been studied in  Refs.~\cite{Nowak:1992um,Nowak:2003ra,Bardeen:1993ae,Bardeen:2003kt}. In Ref. \cite{Ma:2015cfa}, doubly heavy baryons with negative parity were investigated by regarding them as chiral partners of the positive parity heavy baryons. The techniques of SDE were employed in~\cite{Qin:2019hgk} to study the problem involving any number of heavy
quarks. In this article, we extend the work reported in~\cite{Gutierrez-Guerrero:2019uwa} to calculate the masses of
negative parity baryons by employing solutions of the BSE and FE. It naturally requires the study of axial and scalar mesons as well their
corresponding diquarks.
We use a realistic description of baryons as bound states of quarks and non-pointlike and dynamical diquark correlations~\cite{Bashir:2012fs,Barabanov:2020jvn}.
This reduces the problem to a simplified sub-structure of several two-body correlations. The same interaction
which describes mesons also generates corresponding diquark correlations in the colour-antitriplet ($\overline{3}$) channel
with half the attractive strength. The diquarks being a colored correlation are confined within the baryons.
The baryon system is bound due to the interaction between the quarks forming a diquark and the spectator quark
which continuously interchange their roles. The validity of the quark-diquark picture
was confirmed in~\cite{Eichmann:2009qa} at the level of 5\% for the nucleon mass.

Our article has been organized as follows. In a brief Sec.~II, we summarize the main features of the
CI along with the sets of parameters we shall employ in our subsequent analysis. Sec.~III is devoted
to a detailed study of the BSE for all the relevant mesons and the corresponding diquarks. In Sec.~IV,
we solve the FE and calculate the masses of all the spin-1/2 and spin-3/2 baryon parity partners
containing $\tu,\td,\ts,\tc,\tb$ quarks. Sec.~V concludes our work. The mathematical details of our
analysis have been presented in appendices A,B and C.

\section{ Contact interaction: features }
\label{CI-1}

The gap equation for the quarks naturally requires modelling the gluon propagator and the quark-gluon vertex.
In this section, we recall the main truncations and characteristics which define the CI~\cite{GutierrezGuerrero:2010md,Roberts:2010rn,Roberts:2011wy,Roberts:2011cf}~:
\begin{itemize}
\item  The gluon propagator is defined to be independent of any running momentum scale:
\begin{eqnarray}
\label{eqn:contact_interaction}
g^{2}D_{\mu \nu}(k)&=&4\pi\rmh\delta_{\mu \nu} \equiv
\frac{1}{m_{G}^{2}}\delta_{\mu\nu},
\end{eqnarray}
\noindent with $\rmh=\alpha_{\mathrm {IR}}/m_g^2$, $m_g=500\,\MeV$ is a gluon mass scale generated dynamically in QCD~\cite{Boucaud:2011ug,Aguilar:2017dco,Binosi:2017rwj,Gao:2017uox}, and $\alpha_{\mathrm{IR}}$ can be interpreted as the interaction strength in the infrared~\cite{Binosi:2016nme,Deur:2016tte,Rodriguez-Quintero:2018wma}.
\item At  leading-order, the quark-gluon vertex is
\begin{equation}
\Gamma_{\nu}(p,q) =\gamma_{\nu} .
\end{equation}
\item With this kernel, the dressed-quark propagator for a quark of flavor $f$ becomes
\begin{eqnarray}
 \nn && S_f^{-1}(p) = \\
&&  i \gamma \cdot p + m_f +  \frac{16\pi}{3}\rmh \int\!\frac{d^4 q}{(2\pi)^4} \,
\gamma_{\mu} \, S_f(q) \, \gamma_{\mu}\,,\label{gap-1}
\end{eqnarray}
 where $m_f$ is the current-quark mass. The integral possesses quadratic and logarithmic divergences. We choose to regularize them in a Poincar\'e covariant manner. The solution of this equation is~:
\begin{equation}
\label{genS}
S_f^{-1}(p) = i \gamma\cdot p + M_f\,,
\end{equation}
 where $M_f$ in general is the mass function running with a momentum scale. However, within the CI it
 is a constant dressed mass.
\item $M_f$ is determined by
\begin{equation}
M_f = m_f + M_f\frac{4\rmh}{3\pi}\,\,{\cal C}^{\rm iu}(M_f^2)\,,
\label{gapactual}
\end{equation}
where
\bea
\hspace{0.75 cm}
{\cal C}^{\rm iu}(\sigma)/\sigma = \overline{\cal C}^{\rm iu}(\sigma) = \Gamma(-1,\sigma \tau_{\rm uv}^2) - \Gamma(-1,\sigma \tau_{\rm ir}^2),
\eea
 where $\Gamma(\alpha,y)$ is the incomplete gamma-function and $\tau_{\rm ir, uv}$ are respectively, infrared and ultraviolet regulators. A nonzero value for  $\tau_{\mathrm{IR}}\equiv 1/\Lambda_{\mathrm{IR}}$ implements
confinement~\cite{Roberts:2007ji}. Since the CI is a nonrenormalizable theory,
$\tau_{\mathrm{UV}}\equiv 1/\Lambda_{\mathrm{UV}}$ becomes part of the definition of the our model and therefore sets the scale for
all dimensional quantities.
 \end{itemize}
 In this work we report results  using the values in Tables~\ref{parameters},~\ref{table-M}, which correspond to what were dubbed as
 {\em heavy parameters} in~\cite{Gutierrez-Guerrero:2019uwa}. In this choice, the coupling constant and the ultraviolet regulator vary as a function of the quark mass. This behavior was first suggested in~\cite{Bedolla:2015mpa} and later adopted in several subsequent works~\cite{Bedolla:2016yxq,Raya:2017ggu,Gutierrez-Guerrero:2019uwa,Yin:2019bxe,Yin:2021uom}.
 If one wants to go beyond predicting the masses of the hadrons and construct a model which can also predict
charge radii and decay constants, then the study of the heavy sector requires a change in the model parameters with respect to those of the light sector: an increase in the ultraviolet regulator, and a reduction in the coupling strength. Following Ref.~\cite{Raya:2017ggu}, guided by~\cite{Farias:2005cr,Farias:2006cs}, we define a dimensionless coupling $\hat{\alpha}$~:
\begin{equation}
\hat{\alpha}(\Lambda_{\mathrm{UV}})=\rmh\Lambda_{\mathrm{UV}}^2.\label{eqn:dimensionless_alpha}
\end{equation}
In close analogy with the running coupling of QCD with the momentum scale on which it is measured, an inverse logarithmic curve can fit the functional dependence of ${\hat{\alpha}}(\Lambda_{\mathrm {UV}})$ reasonably well. The fit reads:
\begin{equation}
\label{eqn:logaritmicfit} \hat{\alpha}(\Lambda_{\mathrm{UV}})=a\ln^{-1}\left(\Lambda_{\mathrm {UV}}/\Lambda_0\right) \,,
\end{equation}
where $a=0.92$ and $\Lambda_0=0.36$ GeV, see Ref.~\cite{Raya:2017ggu}. With this fit, we can estimate the
value of the coupling strength $\hat{\alpha}(\Lambda_{\mathrm{UV}})$ once a value of $\Lambda_{\mathrm {UV}}$
is assigned. Note that
\begin{eqnarray}
   \Lambda_{\rm IR} = 0.24 {\rm GeV} &\rightarrow& {\rm Confinement \; scale} \nn  \\
   m_g =0.5 {\rm GeV} &\rightarrow& {\rm Infrared \; gluon \; mass \; scale}  \nn \\
   \Lambda_{\rm UV} = 0.905 {\rm GeV} &\rightarrow& {\rm Hadronic \; scale \; for \; light \; hadrons} \nn
\end{eqnarray}
These reproduce the value of the chiral quark condensate accurately. For increasing quark mass involved, $\Lambda_{\rm UV}$ would change.
The chosen strength of the coupling $\hat{\alpha}$ and $\Lambda_{\rm UV}$, which follow the logarithmic curve mentioned above, (along with the current quark masses) are fitted to the lightest pseudo-scalar meson mass and its charge-radius.

 \begin{table}[H]
 \caption{\label{parameters} Ultraviolet regulator and $m_G$ (in GeV) as well as dimensionless coupling constant for different combinations of quarks in a hadron.  $\alpha_{\mathrm {IR}}=\alpha_{\mathrm{IRL}}/Z_H$ with $\alpha_{\mathrm {IRL}}=1.14$, extracted from a best-fit to data, as explained in~\cite{Raya:2017ggu}. Fixed parameters are the gluon mass $m_g =500$ MeV  reported in~\cite{Gao:2017uox} and $\Lambda_{\rm IR} = 0.24$ GeV.}
\begin{center}
\label{parameters1}
\begin{tabular}{@{\extracolsep{0.5 cm}}lccc}
\hline \hline
 quarks & $Z_H$ & $\Lambda_{\mathrm {UV}}\;[\GeV]$ & $m_G$  \\
 \rule{0ex}{3.5ex}
$\tu,\td,\ts$ & 1 & 0.905& 0.165\\
\rule{0ex}{3.5ex}
$\tc,\td,\ts$ & 3.034 & 1.322&0.287 \\
\rule{0ex}{3.5ex}
$\tc$     & 13.122 & 2.305&0.598 \\
\rule{0ex}{3.5ex}
$\tb,\tu,\ts$ & 16.473 & 2.522 & 0.669 \\
\rule{0ex}{3.5ex}
$\tb,\tc$   & 59.056 & 4.131 &1.268 \\
\rule{0ex}{3.5ex}
$\tb$     & 165.848 & 6.559 & 2.125\\
\hline \hline
\end{tabular}
\end{center}
\end{table}
 %
%
%
Table~\ref{table-M} presents current quark masses used herein and dressed masses of $\tu$, $\ts$, $\tc$ and $\tb$ computed from Eq.~(\ref{gapactual}).
\begin{table}[H]
\caption{\label{table-M}
Current ($m_{u,\cdots}$) and dressed masses
($M_{u,\cdots}$) for quarks (GeV), required as input for the BSE and the FE.}
\vspace{0.3cm}
\begin{tabular}{@{\extracolsep{0.2 cm}}cccc}
\hline
\hline
 $m_{\tu}=0.007$ &$m_{\ts}=0.17$ & $m_{\tc}=1.08$ & $m_{\tb}=3.92$   \\
 $\hspace{-.2cm} M_{\tu}=0.367$ & $ \hspace{0.1cm} M_{\ts}=0.53$\; &  $ \hspace{0.0 cm} M_{\tc}=1.52$ & $ \hspace{0.0cm} M_{\tb}=4.68$  \\
 \hline
 \hline
\end{tabular}
\end{table}
The simplicity of the CI allows one to readily compute hadronic observables, such as masses, decay constants, charge radii and form factors. The study of heavy, heavy-light and light meson masses provides a way to determine the masses associated with diquark correlations and those of heavy, heavy-light, and light baryons and their negative parity partners. With this in mind, in the next section, we describe and solve the BSE for mesons and diquarks.
\section{Bethe Salpeter Equation}
\label{CI-1}
The bound-state problem for hadrons characterized by two valence-fermions is studied using the
homogeneous BSE. This equation is~\cite{Salpeter:1951sz}
%
\begin{equation}
[\Gamma(k;P)]_{tu} = \int \! \frac{d^4q}{(2\pi)^4} [\chi(q;P)]_{sr} K_{tu}^{rs}(q,k;P)\,,
\label{genbse}
\end{equation}
where $\Gamma$ is the bound-state's Bethe-Salpeter amplitude (BSA); $\chi(q;P) = S(q+P)\Gamma S(q)$ is the BS wave-function; $r,s,t,u$ represent colour, flavor and spinor indices; and $K$ is the relevant fermion-fermion scattering kernel.  This equation possesses solutions on that discrete set of $P^2$-values for which bound-states exist.
We use the notation introduced in~\cite{Chen:2012qr}, [$\fd,\fdu$] for scalar and pseudoscalar diquarks, and $(\{\fd,\fd\}),(\{\fd,\fdu\})$ for axial-vector and vector diquarks. We describe the details of the meson mass calculation in the following subsection.
\vspace{-3cm}
\begin{figure}[htbp]
\begin{center}
       \includegraphics[scale=0.4,angle=0]{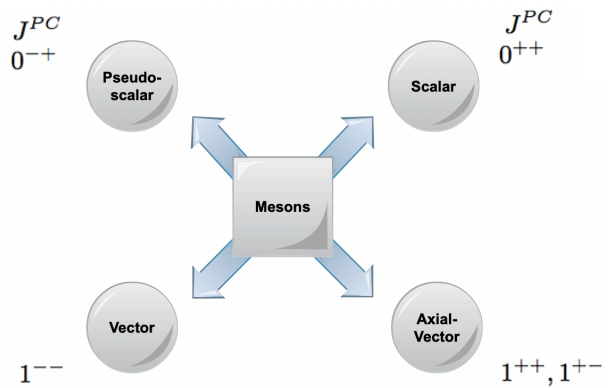}
       \vspace{-3.0cm}
       \caption{\label{cla-me} We study the scalar, pseudoscalar, vector and axial-vector mesons as well as their corresponding diquarks.}
       \end{center}
\end{figure}
\vspace{-1cm}
\subsection{Mesons}
Mesons are classified into groups according to their  total angular momentum (J), parity (P) and the charge-parity (C), employing the usual notation $J^{PC}$. In Fig.~\ref{cla-me}, we show the mesons we study in this work.
 \begin{table}[htbp]
 \caption{\label{ff-BSE-1} Here we list BSA for mesons and the canonical normalization ${\cal N}$. ${\cal K}_{PS},{\cal K}_{V},{\cal K}_{AV},{\cal K}_{S}$ are given in Eqs.~(\ref{kps}) and~(\ref{Kse}), respectively. $M_R = M_{\fd} M_{\fu}/[M_{\fd} + M_{\fu}]$.}
\begin{center}
\label{parameters1}
\begin{tabular}{@{\extracolsep{0.5 cm}}lccc}
\hline \hline
BSA & A & B & ${\cal N}$\\
\rule{0ex}{3.5ex}
  $ \Gamma_{PS}$ & $i\gamma_5$ & $\frac{1}{2M_R}\gamma_5 \gamma\cdot P$ & $6\frac{d {\cal K}_{PS}(Q^2,z)}{dz}\bigg|_{Q^2=z}$ \\ \\
\rule{0ex}{3.5ex}
$ \Gamma_{V,\mu}$ & $\gamma_{\mu}^{T}$ & $\frac{1}{2M_R}\sigma_{\mu\nu}P_\nu$ & $9m_G^2 E_{V}^2\frac{d {\cal K}_{V}(z)}{dz}$  \\ \\
\rule{0ex}{3.5ex}
$\Gamma_{S}$ &$I_D$ &--&$-\frac{9}{2}m_G^2 E_S^2\frac{d {\cal K}_S(z)}{dz}$  \\ \\
\rule{0ex}{3.5ex}
$\Gamma_{AV,\mu}$ & $\gamma_5\gamma_{\mu}^{T}$ &$\gamma_5\frac{1}{2M_R}\sigma_{\mu\nu}P_\nu$ & $-9m_G^2 E_{AV}^2\frac{d {\cal K}_{AV}(z)}{dz}$  \\ \\
\hline \hline
\end{tabular}
\end{center}
\vspace{-0.5cm}
\end{table}
\vspace{0cm}
In a symmetry-preserving treatment using relativistic quantum field theory, pseudoscalar-scalar and vector-axial mesons are chiral partners.
Meson chiral partners are the simplest bound states that are addressed in this work.
%
A general decomposition of the bound-state's BSA for mesons in the CI has the form
\bea
\label{BSA-mesones}
\Gamma_H=A_HE_H +B_H F_H\;,
\eea
where $H={PS,V,AV,S}$ denotes pseudoscalar (PS), vector (V), axial-vector (AV) and scalar (S) mesons, respectively. The explicit form of the BSA canonical normalization for different types of mesons is displayed in Table~\ref{ff-BSE-1}. We
adopt the notation
\bea
      \frac{d {\cal K}_{PS}(Q^2,z)}{dz}\bigg|_{Q^2=z} \equiv \frac{d {\cal K}_{PS}(z)}{dz}\bigg|_{z} \nn \,.
\eea
The BSE for pseudoscalar meson comprised of quark with flavour $\fd$ and antiquark with flavour $\fu$ is
\bea
\nn\label{kps}{\cal K}_{PS}(Q,P)&=& {\rm tr}_{\rm D} \int\! \frac{d^4q}{(2\pi)^4} \Gamma_{\Meps}(-Q)
 \frac{\partial}{\partial P_\mu} S_{\fd}(q+P)
 \\
 &\times&\Gamma_{\Meps}(Q)\, S_{\fu}(q)\,,
\eea
where P is the total momentum of the bound state. The explicit matrix form of the BSE is:
\begin{equation}
\label{bsefinalEf}
\left[
\begin{array}{c}
E_{\Meps}(P)\\
\rule{0ex}{3.0ex}
F_{\Meps}(P)
\end{array}
\right]
= \frac{4 \rmh}{3\pi}
\left[
\begin{array}{cc}
{\cal K}_{EE}^{\Meps} & {\cal K}_{EF}^{\Meps} \\
\rule{0ex}{3.0ex}
{\cal K}_{FE}^{\Meps}& {\cal K}_{FF}^{\Meps}
\end{array}\right]
\left[\begin{array}{c}
E_{\Meps}(P)\\
\rule{0ex}{3.0ex}
F_{\Meps}(P)
\end{array}
\right],
\end{equation}
with $\rmh=\alpha_{\mathrm {IR}}/m_g^2$ and
\begin{subequations}
\label{pionKernel}
\begin{eqnarray}
\nonumber
\nn {\cal K}_{EE}^{\Meps} &=&
\int_0^1d\alpha \bigg\{
{\cal C}^{\rm iu}(\omega^{(1)})  \\
\nn&&+ \bigg[ M_{\fu} M_{\fd}-\alpha (1-\alpha) P^2 - \omega^{(1)}\bigg]
\, \overline{\cal C}^{\rm iu}_1(\omega^{(1)})\bigg\},\\
\nn {\cal K}_{EF}^{\Meps} &=& \frac{P^2}{2 M_R} \int_0^1d\alpha\, \bigg[(1-\alpha)M_{\fu}+\alpha M_{\fd}\bigg]\overline{\cal C}^{\rm iu}_1(\omega^{(1)}),\\
\nn{\cal K}_{FE}^{\Meps} &=& \frac{2 M_R^2}{P^2} {\cal K}_{EF}^K ,\\
\nn {\cal K}_{FF}^{\Meps} &=& - \frac{1}{2} \int_0^1d\alpha\, \bigg[ M_{\fu} M_{\fd}+(1-\alpha) M_{\fu}^2+\alpha M_{\fd}^2\bigg] \\
\nn &\times &\overline{\cal C}^{\rm iu}_1(\omega^{(1)})\,,
\end{eqnarray}
\end{subequations}
where $\alpha$ is a Feynman parameter and the new functions $\omega^{(1)}\equiv\omega(M_{\fu}^2,M_{\fd}^2,\alpha,P^2)$ and ${\cal C}^{\rm iu}_1(z) $ are
\begin{eqnarray}
\label{eq:omega}
\nn&& \hspace{-0.4cm} \omega^{(1)} =M_{\fu}^2 (1-\alpha) + \alpha M_{\fd}^2 + \alpha(1-\alpha) P^2\,,\\
&& \hspace{-0.4cm} \nn{\cal C}^{\rm iu}_1(z) = - z (d/dz){\cal C}^{\rm iu}(z) = z\left[ \Gamma(0,M^2 \tau_{\rm uv}^2)-\Gamma(0,M^2 \tau_{\rm ir}^2)\right] .\rule{2em}{0ex}
\label{eq:C1}
\end{eqnarray}
The eigenvalue equations for vector, axial-vector and scalar mesons are:
\bea
\label{eig}
 1\hspace{-0.1cm}-\hspace{-0.06cm}{\cal K}_{V}(-m_{V}^2)\hspace{-0.06cm}=\hspace{-0.06cm}1\hspace{-0.06cm}+\hspace{-0.06cm}{\cal K}_{AV}(-m_{AV}^2)\hspace{-0.06cm}=\hspace{-0.06cm} 1\hspace{-0.06cm}+\hspace{-0.06cm}{\cal K}_{S}(-m_{S}^2)\hspace{-0.06cm}=\hspace{-0.06cm}0,
\eea
we have defined
\bea  && \hspace{-4mm} \nn {\cal L}_{\Me}(P^2)= M_{\fu} M_{\fd} - (1-\alpha)M_{\fu}^2-\alpha M_{\fd}^2-2\alpha(1-\alpha)P^2,\\
&& \hspace{-4mm}\nn {\cal L}_{G}(P^2)=M_{\fd} M_{\fu}+\alpha (1-\alpha)P^2,
\eea
and
\begin{eqnarray}\nn
\label{KastKernel}
{\cal K}_{\Me}(P^2)&=& \frac{2\rmh}{3\pi} \int_0^1d\alpha\,
{\cal L}_{\Me}(P^2)
\overline{\cal C}_1^{\rm iu}(\omega^{(1)}),\\ \nn \\
\nn {\cal K}_{AV}(P^2) &=&
\frac{2\rmh}{3\pi } \int_0^1d\alpha\,\bigg[
{\cal C}_1^{\rm iu}(\omega^{(1)})
+ {\cal L}_{G}(P^2) \overline{\cal C}_1^{\rm iu}(\omega^{(1)})\bigg],\\
\nonumber
{\cal K}_{S}(P^2) &=&
- \frac{4\rmh}{3\pi}
\int_0^1d\alpha\,\bigg[-{\cal L}_G
\overline{\cal C}_1^{\rm iu}(\omega^{(1)})\\
 &+&\bigg({\cal C}^{\rm iu}(\omega^{(1)})
-{\cal C}_1^{\rm iu}(\omega^{(1)})\bigg) \bigg], \label{Kse}
\end{eqnarray}
The equations (\ref{bsefinalEf},\ref{eig}) have a solution when $P^2=-m_{H}^2$. Then the eigenvector
corresponds to the BSA of the meson. We consider mesons with five flavors ($\tu$,$\td$,$\ts$,$\tc$,$\tb$).
It has long been known that the rainbow-ladder truncation describes vector meson and flavour-nonsinglet
pseudoscalar meson ground-states very well but fails for their parity partners~\cite{Qin:2011xq,Qin:2011dd,Maris:2006ea,Cloet:2007pi,Chen:2012qr}.
It was found that DCSB generates a large dressed-quark anomalous chromomagnetic moment and consequently the spin-orbit splitting between ground-state mesons and their parity partners is dramatically enhanced~\cite{Bermudez:2017bpx,Bashir:2011dp,Chang:2010hb,Chang:2011ei,Chang:2010jq}. This is the mechanism responsible for a magnified splitting between parity partners; namely, there are essentially nonperturbative DCSB corrections to the rainbow-ladder kernels, which largely cancel in the pseudoscalar and vector channels but add constructively in the scalar and axial-vector channels.
In this connection, we follow~\cite{Roberts:2011cf} and introduce spin-orbit repulsion into the scalar- and pseudovector meson channels through the artifice of a phenomenological coupling $g_{SO} \leq 1$, introduced as a single, common factor multiplying the kernels defined in Eqs.~(\ref{bsefinalEf}),~(\ref{eig}).
$g_{SO}$ mimics the dressed-quark chromomagnetic moment
in full QCD.
The  first numerical value of $g_{SO} = 0.24$ was introduced in ~\cite{Roberts:2011cf} and later refined in Refs. \cite{Yin:2021uom,Lu:2017cln,Gutierrez-Guerrero:2019uwa}. For mesons with $J^P=0^+\;,1^+$ we use
\bea
g_{SO}^{0^+}=0.32\;,\;\;\;g_{SO}^{1^+}=0.25\;.
\eea
The value $g_{SO}^{1^+}=0.25$ in the axial vector channel guarantees the effect of spin-orbit repulsion and reproduces the desirable experimental value for the $a_1-\rho$ mass-splitting \cite{Yin:2021uom,Lu:2017cln,Gutierrez-Guerrero:2019uwa}.
On the other hand, $g_{SO}^{0^+}=0.32$ is chosen
to produce a mass difference of approximately $0.3$ GeV
between the quark-core of the $0{^+}(\tu\bar\td)$ (which we call  $\sigma(\tu\bar\td)$) and that of the $\rho$-meson (as obtained with beyond-RL kernels). The choice of
$g_{SO}= 1$  indicates no repulsion and no additional interaction beyond that generated by the rainbow-ladder kernel.
The numerical results for the pseudoscalar, and scalar mesons are reported in Tables~\ref{table-mesones-pseudo},\ref{table-mesones-esc}.
For pseudoscalar mesons, the computed masses in Table \ref{table-mesones-pseudo} are also compared
with experimental values  and the largest percentage difference is for pion (about 0.7\%), which becomes smaller for heavy-light mesons and is zero for $\eta_c$ and $\eta_b$.

\begin{table}[htbp]
\caption{\label{table-mesones-pseudo}
Computed masses for pseudoscalar mesons (GeV) and BSAs with the parameters of Tables~\ref{parameters},~\ref{table-M}. }
\begin{center}
\begin{tabular}{@{\extracolsep{0.2 cm}}cccccc}
\hline
\hline
Mesons   & Exp. & CI & E & F & Charge \\
\rule{0ex}{2.5ex}
$\pi(\tu\bar{\td})$ &0.139&0.14&3.60&0.47&1\\
\rule{0ex}{2.5ex}
$K(\tu\bar{\ts})$ &0.493&0.49&3.81&0.59&1\\
\rule{0ex}{2.5ex}
$h_s(\ts\bar{\ts})$ &--& 0.69&4.04&0.74&0\\
\rule{0ex}{2.5ex}
$D^{0}(\tc\bar{\tu})$ &1.86& 1.87&3.03&0.37&0\\
\rule{0ex}{2.5ex}
$D^{+}_{\ts}(\tc\bar{\ts})$ &1.97 & 1.96& 3.24 & 0.51&1 \\
\rule{0ex}{2.5ex}
$B^{+} (\tu\bar{\tb})$ &5.28&5.28& 1.50&0.09&1 \\
\rule{0ex}{2.5ex}
$B_s^0(\ts\bar{\tb})$  &5.37& 5.37&1.59&0.13&0\\
\rule{0ex}{2.5ex}
$B_{\tc}^{+}(\tc\bar{\tb})$ &6.27&6.29&0.73& 0.11&1 \\
\rule{0ex}{2.5ex}
$\eta_c(\tc\bar{\tc})$ &2.98& 2.98&2.16& 0.41&0\\
\rule{0ex}{2.5ex}
$\eta_{\tb}(\tb\bar{\tb})$ &9.40& 9.40&0.48&0.10&0\\
\hline
\hline
\end{tabular}
\end{center}
\end{table}
Table \ref{table-mesones-esc} depicts the scalar mesons.
In the scalar channel, the highest percentage difference is for $K_0^*$ and has a value of 7\%.
\begin{table}[htbp]
\caption{\label{table-mesones-esc}
Computed masses for scalar mesons (GeV) and BSA with the parameters listed in Tables~\ref{parameters},~\ref{table-M} and $g_{SO}=0.32$ }
\begin{center}
\begin{tabular}{@{\extracolsep{0.5 cm}}cccc}
\hline
\hline
Mesons   &Exp.& CI & E\\
\rule{0ex}{2.5ex}
$\sigma(\tu\bar{\td})$ & 1.2 &1.22&0.66\\
\rule{0ex}{2.5ex}
$K_0^*(\tu\bar{\ts})$ &1.430&1.33&0.65\\
\rule{0ex}{2.5ex}
$f_0(\ts\bar{\ts})$ &--&1.34&0.64\\
\rule{0ex}{2.5ex}
$D_0^*(\tc\bar{\tu})$&2.300&2.32&0.39\\
\rule{0ex}{2.5ex}
$D_{s0}^*(\tc\bar{\ts})$& 2.317&2.43&0.37\\
\rule{0ex}{2.5ex}
$B_{0}^*(\tu\bar{\tb})$& --&5.50&0.21\\
\rule{0ex}{2.5ex}
$B_{s0}(\ts\bar{\tb})$&--&5.59&0.20\\
\rule{0ex}{2.5ex}
$B_{c0}(\tc\bar{\tb})$&--&6.45&0.08\\
\rule{0ex}{2.5ex}
$\chi_{c0}(\tc\bar{\tc})$&3.414&3.35&0.16\\
\rule{0ex}{2.5ex}
$\chi_{b0}(\tb\bar{\tb})$ &9.859&9.50&0.04\\
\hline
\hline
\end{tabular}
\end{center}
\end{table}
 The analysis of $\pi(\tu\bar\td)$ and $\sigma(\tu\bar\td)$ masses indicates a difference of approximately $1.061$ GeV. However, this difference is less pronounced for mesons composed of two heavy quarks for example $\eta_b(\tb\bar{\tb})$ and  $\chi_{b0}(\tb\bar{\tb})$ which have very close masses in our model.
  Pseudoscalar and scalar  meson must satisfy the following mass relation:
 \bea\label{eq:1}\nn
    m_{D_{\ts}^{+}(\tc\overline{\ts})} - m_{D^{0}(\tc\overline{\tu})} + m_{B^{+}(\tu\overline{\tb})} - m_{B_{\ts}^{0}(\ts\overline{\tb})} &=& 0,\\
    m_{D_{\ts0}^{*}(\tc\overline{\ts})} - m_{D^{*}_{0}(\tc\overline{\tu})} + m_{B^{*}_{0}(\tu\overline{\tb})} - m_{B_{\ts0}(\ts\overline{\tb})} &=& 0.
 \eea
In our model, these equations are exactly satisfied for pseudoscalar mesons and deviate only up to 2\% for scalars.
Vector and axial vector mesons are reported in Tables~\ref{table-mesones-vec},~\ref{table-mesones-av}.
It is straightforward to observe that the $\rho$-meson has the greatest difference as compared to the empirical value (approximately 19\%). Note that this value had already been used in several previous works using this model \cite{Roberts:2011wy,Chen:2012qr,Roberts:2011cf,Yin:2019bxe,Yin:2021uom,Gutierrez-Guerrero:2019uwa}.
Although this value is larger than expected, using the corresponding parameters we can calculate decay constants which lie very close to the experimental ones. Furthermore, when the masses of the corresponding diquarks are calculated to predict the masses of baryons, they closely resemble the experimental value.
As in the previous cases, the percentage differences between our results and the experimental ones are smaller when heavy quarks
are included which constitute the gist of our
study.

\begin{figure}[htbp]
       \centerline{
       \includegraphics[scale=0.3,angle=-90]{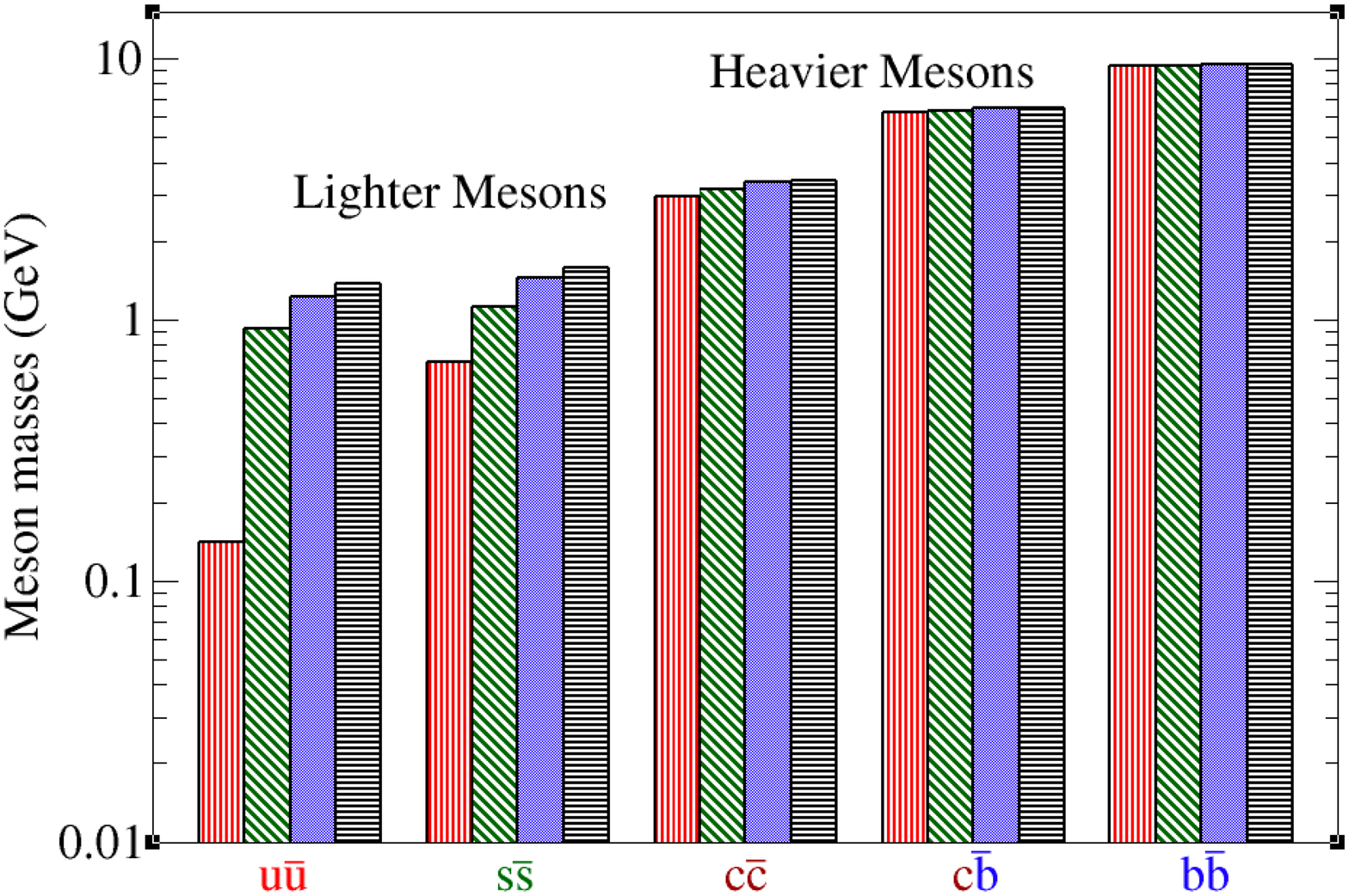}}
       \caption{ \label{Fig1} The light and heavy meson masses (GeV): red (vertical lines)-pseudoscalar mesons, green (diagonal lines)-vector mesons, blue (solid filled rectangles)-scalar mesons and black (horizontal lines)-axial-vector mesons.}
       \vspace{-1cm}
\end{figure}

\begin{figure}[htbp]
       \centerline{
       \includegraphics[scale=0.3,angle=-90]{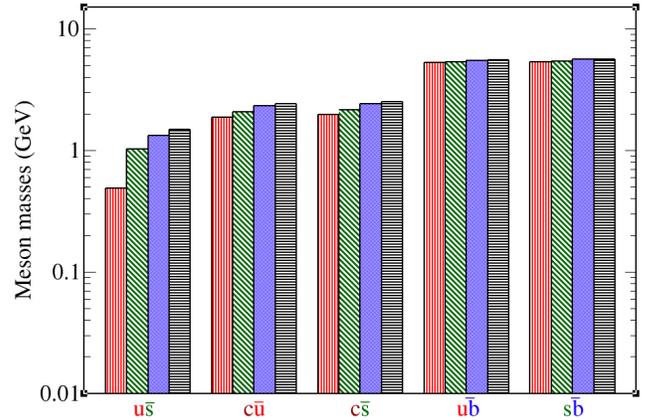}}
       \caption{ \label{Fig2} The heavy-light meson masses (GeV): red (vertical lines)-pseudoscalar mesons, green (diagonal lines)-vector mesons, blue (solid filled rectangles)-scalar mesons and black (horizontal lines)-axial-vector mesons.}
\end{figure}

For the case of axial vector channel, mean-absolute-relative-difference between the entries in columns 3 and 4 of Table \ref{table-mesones-av} does not exceed 12\% and for $\chi_{c1}$ and $\chi_{b1}$ it is between 3\% and 4\%
The mass-splitting between the opposite parity partners $\rho(\tu\bar\td)-a_1(\tu\bar\td)$ is $440$ MeV and $100$ MeV for $\Upsilon(\tb\bar{\tb})$-$\chi_{b1}(\tb\bar{\tb})$ (less than 5\% error in both cases). All computed values for ground-state heavy-light mesons exhibit a similar mass difference compared to their chiral partners, which decreases with increasing meson mass.
 \begin{table}[htbp]
\caption{\label{table-mesones-vec}
Vector meson masses (GeV) and BSA computed with the parameters listed in Tables~\ref{parameters},~\ref{table-M}.} 
\vspace{0.0cm}
\begin{center}
\begin{tabular}{@{\extracolsep{0.2 cm}}cccc}
\hline
\hline
Mesons  & Exp.& CI & E \\
\rule{0ex}{2.5ex}
$\rho$ $(\tu\bar{\td})$ & 0.78 & 0.93 & 1.53\\
\rule{0ex}{2.5ex}
$K_1$ $(\tu\bar{\ts})$ & 0.89 & 1.03 & 1.62\\
\rule{0ex}{2.5ex}
$\phi$ $(\ts\bar{\ts})$ & 1.02 & 1.12 & 1.73\\
\rule{0ex}{2.5ex}
$D^{*0}(\tc\bar{\tu})$& 2.01 & 2.06 & 1.23  \\
\rule{0ex}{2.5ex}
$D_{\ts}^{*}(\tc\bar{\ts})$ & 2.11& 2.14 &1.32 \\
\rule{0ex}{2.5ex}
$B^{+*}(\tu\bar{\tb})$ &5.33& 5.33 & 0.65 \\
\rule{0ex}{2.5ex}
$B_{\ts}^{0*}(\ts\bar{\tb})$ & 5.42 & 5.41 & 0.67\\
\rule{0ex}{2.5ex}
$B_{\tc}^{*}(\tc\bar{\tb})$  &--& 6.32 & 0.27 \\
\rule{0ex}{2.5ex}
$\Jpsi$ $(\tc\bar{\tc})$ & 3.10& 3.15 & 0.61 \\
\rule{0ex}{2.5ex}
$\Upsilon(\tb\bar{\tb})$ &9.46& 9.42 &0.15 \\
\hline
\hline
\end{tabular}
\end{center}
\end{table}
\begin{table}[htbp]
\caption{\label{table-mesones-av}
Axial-vector meson masses (GeV) and BSA, computed with the parameters listed in Tables~\ref{parameters},~\ref{table-M} and $g_{SO}=0.25.$} 
\begin{center}
\begin{tabular}{@{\extracolsep{0.2 cm}}cccc}
\hline
\hline
Mesons   & Exp. & CI & E \\
\rule{0ex}{2.5ex}
$a_1(\tu\bar{\td})$ & 1.260 &1.37&0.32\\
\rule{0ex}{2.5ex}
$K_1(\tu\bar{\ts})$ &1.34&1.48&0.32\\
\rule{0ex}{2.5ex}
$f_1(\ts\bar{\ts})$ &1.43&1.58&0.32\\
\rule{0ex}{2.5ex}
$D_1(\tc\bar{\tu})$&2.420&2.41&0.20\\
\rule{0ex}{2.5ex}
$D_{s1}(\tc\bar{\ts})$&2.460&2.51&0.19\\
\rule{0ex}{2.5ex}
$B_{1}(\tu\bar{\tb})$&5.721&5.55&0.11\\
\rule{0ex}{2.5ex}
$B_{s1}(\ts\bar{\tb})$&5.830&5.64&0.10\\
\rule{0ex}{2.5ex}
$B_{cb}(\tc\bar{\tb})$&--&6.48&0.04\\
\rule{0ex}{2.5ex}
$\chi_{c1}(\tc\bar{\tc})$&3.510&3.40&0.08\\
\rule{0ex}{2.5ex}
$\chi_{b1}(\tb\bar{\tb})$ &9.892&9.52&0.02\\
\hline
\hline
\end{tabular}
\end{center}
\end{table}
Moreover, one can immediately see that in all cases the pseudoscalar are the lightest and  axial-vector the heaviest mesons. This information is represented pictorially in Figs.~\ref{Fig1} and \ref{Fig2}.
The same behavior is observed for light, heavy and heavy-light mesons. However, it is more conspicuous for those that are composed of two light quarks. The calculation of baryons with negative parity requires the masses and amplitudes of the diquarks $J^P=0^+\;,1^+,\;0^-$ and $1^-$. It is for this reason that in this section we have included the axial and scalar mesons. The equal spacing rules for vector and axial vector mesons are:
\bea\label{eq:2} \nn
    m_{D_{\ts}^{*}(\tc\overline{\ts})} - m_{D^{0*}(\tc\overline{\tu})} + m_{B^{+*}(\tu\overline{\tb})} - \nn m_{B_{\ts}^{0*}(\ts\overline{\tb})} &=& 0,\\ \nn
    m_{D_{\ts1}(\tc\overline{\ts})} - m_{D_1(\tc\overline{\tu})} + m_{B_1(\tu\overline{\tb})} - m_{B_{\ts1}(\ts\overline{\tb})} &=& 0.\\
 \eea
Using the results obtained in Tables~\ref{table-mesones-vec},~\ref{table-mesones-av} we instantly infer that the equations \ref{eq:2} are satisfied identically for vector mesons while the deviation for axial vector mesons is less than 1\%.
With these results it is immediate to verify the following equations
\bea \label{gmo-1}
&&\hspace{-0.7cm}m_{B_c^{*}(\tc\bar{\tb})}-m_{B_s^{0*}(\ts\bar{\tb})}-m_{B_c^{+}(\tc\bar{\tb})}+m_{B_s^{0}(\ts\bar{\tb})}\approx 0\,,\\
\label{gmo-2}&& \hspace{-0.7cm}m_{B_{\ts}^{0*}(\ts\bar{\tb})}-m_{B^{+*}(\tu\bar{\tb})}-m_{B_{\ts}^0(\ts\bar{\tb})}+m_{B^{+}(\tu\bar{\tb})}= 0\,,\\
\label{gmo-3}&&\hspace{-0.7cm}m_{B_{\ts}^{0*}(\ts\bar{\tb})}-m_{B^{+*}(\tu\bar{\tb})}-m_{D_{\ts}^{+}(\tc\bar{\ts})}+m_{D^{0}(\tc\bar{\tu})}=0\,, \\
\label{gmo-4}&&\hspace{-0.7cm}m_{\eta_{\tb}(\tb\bar{\tb})}-m_{\eta_{\tc}(\tc\bar{\tc})}-2m_{B_{\ts}^{0*}(\ts\bar{\tb})}+2m_{D_{\ts}^{*}(\tc\bar{\ts})}\approx 0\,,\\
\label{gmo-5}&&\hspace{-0.7cm}m_{\eta_{\tb}(\tb\bar{\tb})}-m_{\eta_{\tc}(\tc\bar{\tc})}-2m_{B_{\ts}^0(\ts\bar{\tb})}+2m_{D_{\ts}^{+}(\tc\bar{\ts})}=0\,,\\
\label{gmo-6}&&\hspace{-0.7cm}m_{B_{\ts}^{0*}(\ts\bar{\tb})}-m_{D_{\ts}^{*}(\tc\bar{\ts})}-m_{B_{\ts}^0(\ts\bar{\tb})}+m_{D_{\ts}^{+}(\tc\bar{\ts})}=0\,.\\
\label{gmo-7}&&\hspace{-0.7cm}m_{\Upsilon(\tb\bar{\tb}) }-m_{\Jpsi(\tc\bar{\tc})}-2m_{B_{\ts}^0(\ts\bar{\tb})}+2m_{D_{\ts}^{+}(\tc\bar{\ts})}=0\,.\\
\label{gmo-8}&&\hspace{-0.7cm}m_{\Upsilon(\tb\bar{\tb}) }-m_{\Jpsi(\tc\bar{\tc})}-m_{\eta_{\tb}(\tb\bar{\tb})}+m_{\eta_{\tc}(\tc\bar{\tc})}\approx 0\,.\\
\label{gmo-9}&&\hspace{-0.7cm}m_{\Upsilon(\tb\bar{\tb}) }-m_{\Jpsi(\tc\bar{\tc})}-2m_{B_{\ts}^{0*}(\ts\bar{\tb})}+2m_{D_{\ts}^{*}(\tc\bar{\ts})}\approx 0\,.
\eea
We test these mass relations, Eqs.~(\ref{gmo-1}-\ref{gmo-9}), against experiment. The deviation from these mass relations is listed in Table~\ref{tab3:Resultados-reglas-espaciados-masas}.
 \begin{table}[htbp]
\caption{The deviation from equal spacing rules for the meson masses, Eqs.~(\ref{gmo-1}-\ref{gmo-9}) is given in GeV, both for
the experimental results and the computed masses from the CI.}
    \label{tab3:Resultados-reglas-espaciados-masas}
    \centering
        \begin{tabular}{ c | c | c | c | c}
            \hline\hline
             & (CI,+) &(Exp.,+)  & (CI,-)& (Exp.,-)   \\
            \hline\hline
            Eq. (\ref{gmo-1}) & -0.01 & $\dots$ & 0.02 & $\dots$ \\
            \rule{0ex}{2.5ex}
            Eq. (\ref{gmo-2}) & -0.01 & 0.0 & 0 & $\dots$ \\
            Eq. (\ref{gmo-3}) & -0.01 & -0.02 & -0.02 & 0.092 \\
            \rule{0ex}{2.5ex}
            Eq. (\ref{gmo-4}) & -0.12 & -0.2  & -0.11 & -0.295 \\
            \rule{0ex}{2.5ex}
            Eq. (\ref{gmo-5}) & -0.4 & -0.38  & -0.17 & $\cdots$ \\
            \rule{0ex}{2.5ex}
            Eq. (\ref{gmo-6}) & -0.14 & -0.09 & -0.03 & $\cdots$ \\
            \rule{0ex}{2.5ex}
            Eq. (\ref{gmo-7}) & -0.55 & -0.44& -0.2 & $\cdots$ \\
            \rule{0ex}{2.5ex}
            Eq. (\ref{gmo-8}) & -0.15 & -0.06  & -0.03 & -0.063\\
            \rule{0ex}{2.5ex}
            Eq. (\ref{gmo-9}) & -0.27 & -0.26 &-0.14 & -0.358  \\
            \hline\hline
        \end{tabular}
\end{table}


\subsection{Diquarks}
Once we have studied the masses of the mesons, the calculation of the diquarks is immediate.
In the notation for diquarks, $H=DS,DAV,DPS,DV$ correspond to scalar, axial-vector, pseudoscalar and vector diquarks, respectively. The BSA for diquarks has the same form as Eq.~(\ref{BSA-mesones}). The explicit coefficients and the normalization conditions are shown in Table~\ref{BSA-diquarks}.
The colour factor for mesons and diquarks is different owing to the fact that diquarks are color antitriplets, not singlets. The canonical normalization condition for diquarks and mesons is almost identical, with the only difference being the replacement ${\cal N}_C=3 \to 2$.
 \begin{table}[htbp]
 \caption{\label{ff-BSE} Here we list BSA for diquarks and the canonical normalization ${\cal N}$.}
\begin{center}
\label{BSA-diquarks}
\begin{tabular}{@{\extracolsep{0.5 cm}}lccc}
\hline \hline
BSA & A & B & ${\cal N}$\\
\rule{0ex}{3.5ex}
$\Gamma_{DS}$  &$i\gamma_5$&$\frac{1}{2M_R}\gamma_5 \gamma\cdot P$& $4\frac{d {\cal K}_{PS}(Q^2,z)}{dz}\bigg|_{Q^2=z}$ \\
\rule{0ex}{3.5ex}
$\Gamma_{DAV,\mu}$  & $\gamma_{\mu}^{T}$ & $\frac{1}{2M}\sigma_{\mu\nu}P_\nu $ &$ 6m_G^2 E_{DAV}^2\frac{d {\cal K}_V(z)}{dz}$  \\
\rule{0ex}{3.5ex}
$\Gamma_{DPS}$  &$I_D$ &--&$-3m_G^2 E_{DPS}^2\frac{d {\cal K}_S(z)}{dz}$  \\
\rule{0ex}{3.5ex}
$\Gamma_{DV,\mu}$ & $\gamma_5\gamma_{\mu}^{T}$ &$\frac{1}{2M}\sigma_{\mu\nu}P_\nu $&$-6m_G^2 E_{DV}^2\frac{d {\cal K}_{AV}(z)}{dz}$  \\
\hline \hline
\end{tabular}
\end{center}
\end{table}
The eigenvalue equations in the case of diquarks are:
\begin{equation}
\label{bsefinalE}
\left[
\begin{array}{c}
E_{\Ds}(P)\\
\rule{0ex}{3.0ex}
F_{\Ds}(P)
\end{array}
\right]
= \frac{4 \rmh}{6\pi}
\left[
\begin{array}{cc}
{\cal K}_{EE}^{\Meps} & {\cal K}_{EF}^{\Meps} \\
\rule{0ex}{3.0ex}
{\cal K}_{FE}^{\Meps}& {\cal K}_{FF}^{\Meps}
\end{array}\right]
\left[\begin{array}{c}
E_{\Ds}(P)\\
\rule{0ex}{3.0ex}
F_{\Ds}(P)
\end{array}
\right].
\end{equation}
The equations that will give us the masses of the axial-vector, vector and pseudoscalar diquarks are
\bea
\nn 0 & = & 1 - \frac{1}{2}{\cal K}_{\Mv}(-m_{\Dav}^2),\\
\nn0 & = & 1 + \frac{1}{2}{\cal K}_{\Mav}(-m_{\Dv}^2),\\
0 & = & 1 + \frac{1}{2}{\cal K}_{\Ms}(-m_{\Dps}^2).
\label{dq}\eea
From Eqs.~(\ref{bsefinalE}) and~(\ref{dq}) it follows that one may obtain the mass and BSA for a diquark with spin-parity $J^P$ from the equation for a $J^{-P}$ meson in which the only change is halving the interaction strength. The flipping of the sign in parity occurs because fermions and antifermions have opposite parity.
In this truncation, the diquark masses
again correspond to $P^2=-m_H^2$.
We therefore present results for the masses of diquark correlations in Tables~\ref{table-diquarks-scalar},~\ref{table-di-pscalar},~\ref{table-diquarks-Av},~\ref{table-Vector-diquaks}. In the case of pseudoscalar and vector diquarks we have multiplied $g_{SO}$ by a factor 1.8 as in~\cite{Lu:2017cln}.
This modification generates less repulsion. Physically, this might be understood by acknowledging that valence-quarks within a diquark are more loosely correlated than the valence-quark and -antiquark pair in a bound-state meson. Consequently, spin-orbit repulsion in diquarks should be less pronounced than it is in the corresponding mesons.
With the diquark masses and amplitudes described herein one can construct all Faddeev kernels associated with ground-state octet and decouplet baryons, and their chiral partners.

 \begin{table}[htbp]
\caption{\label{table-diquarks-scalar}
Computed masses for scalar diquarks (GeV) and BSA with the parameters in Tables~\ref{parameters},~\ref{table-M}. }
\begin{center}
\begin{tabular}{@{\extracolsep{0.2 cm}}cccc}
\hline
\hline
Diquarks & Mass & E & F \\
\rule{0ex}{2.5ex}
$[\tu\td]_{0^+}$ &0.77&2.74&0.31\\
\rule{0ex}{2.5ex}
$[\tu\ts]_{0^+}$ &0.92&2.88&0.39\\
\rule{0ex}{2.5ex}
$[\ts\ts]_{0^+}$&1.06&3.03&0.50\\
\rule{0ex}{2.5ex}
$[\tc\tu]_{0^+}$&2.08&2.00&0.23\\
\rule{0ex}{2.5ex}
$[\tc\ts]_{0^+}$&2.17&2.11&0.32 \\
\rule{0ex}{2.5ex}
$[\tu\bar{\tb}]_{0^+}$&5.37&0.99&0.06\\
\rule{0ex}{2.5ex}
$[\ts\tb]_{0^+}$ &5.46&1.00&0.08\\
\rule{0ex}{2.5ex}
$[\tc\tb]_{0^+}$ &6.35&0.42&0.07\\
\rule{0ex}{2.5ex}
$[\tc\tc]_{0^+}$ &3.17&0.96&0.19\\
\rule{0ex}{2.5ex}
$[\tb\tb]_{0^+}$ &9.43&0.23&0.05\\
\hline
\hline
\end{tabular}
\end{center}
\end{table}
\begin{table}[htbp]
\caption{\label{table-di-pscalar}
Computed masses for pseudoscalar diquarks (in \GeV) with the parameters in Tables~\ref{parameters},~\ref{table-M}. The expressions with superscript * are obtained with $g_{SO}=0.32*1.8.$}
\begin{center}
\begin{tabular}{@{\extracolsep{0.5 cm}}cccccc}
\hline
\hline
Diquark  & Mass & $E$ & Mass$^*$ & $E^*$   \\
\rule{0ex}{2.5ex}
$[\tu\td]_{0^-}$ &1.30&0.54&1.15&1.06\\
\rule{0ex}{2.5ex}
$[\tu\ts]_{0^-}$ &1.41&0.54&1.27&1.05\\
\rule{0ex}{2.5ex}
$[\ts\ts]_{0^-}$ &1.52&0.53&1.40&1.03\\
\rule{0ex}{2.5ex}
$[\tc\tu]_{0^-}$ &2.37&0.32&2.28&0.64\\
\rule{0ex}{2.5ex}
$[\tc\ts]_{0^-}$ &2.47&0.31&2.40&0.61\\
\rule{0ex}{2.5ex}
$[\tu\tb]_{0^-}$ &5.53&0.18&5.47&0.34\\
\rule{0ex}{2.5ex}
$[\ts\tb]_{0^-}$ &5.62&0.14&5.57&0.32\\
\rule{0ex}{2.5ex}
$[\tc\tb]_{0^-}$ &6.47&0.07&6.44&0.13\\
\rule{0ex}{2.5ex}
$[\tc\tc]_{0^-}$ &3.38&0.14&3.33&0.25\\
\rule{0ex}{2.5ex}
$[\tb\tb]_{0^-}$ &9.51& 0.04&9.50&0.07\\
\hline
\hline
\end{tabular}
\end{center}
\end{table}
 \begin{table}[htbp]
\caption{\label{table-diquarks-Av}
Axial-vector diquarks masses (in \GeV), computed with the parameters listed in Tables~\ref{parameters},~\ref{table-M}.} 
\vspace{0.0cm}
\begin{center}
\begin{tabular}{@{\extracolsep{0.2 cm}}cccc}
\hline
\hline
Diquark & Mass & E \\
\rule{0ex}{2.5ex}
$\{\tu\td\}_{1^+}$ &1.06&1.30\\
\rule{0ex}{2.5ex}
$\{\tu\ts\}_{1^+}$ &1.16 &1.36\\
\rule{0ex}{2.5ex}
$\{\ts\ts\}_{1^+}$ &1.25&1.42\\
\rule{0ex}{2.5ex}
$\{\tc\tu\}_{1^+}$&2.16 &0.93  \\
\rule{0ex}{2.5ex}
$\{\tc\ts\}_{1^+}$& 2.25 &0.95 \\
\rule{0ex}{2.5ex}
$\{\tu\tb\}_{1^+}$ &5.39&0.48\\
\rule{0ex}{2.5ex}
$\{\ts\tb\}_{1^+}$ &5.47&0.48\\
\rule{0ex}{2.5ex}
$\{\tc\tb\}_{1^+}$ &6.35&0.20 \\
\rule{0ex}{2.5ex}
$\{\tc\tc\}_{1^+}$ &3.22&0.41 \\
\rule{0ex}{2.5ex}
$\{\tb\tb\}_{1^+}$&9.44&0.11 \\
\hline
\hline
\end{tabular}
\end{center}
\end{table}
\begin{table}[htbp]
\caption{\label{table-Vector-diquaks}
Vector diquark masses (in \GeV), computed with the parameters listed in Tables~\ref{parameters},~\ref{table-M}. The expressions with the superscript * are obtained with $g_{SO}=0.25*1.8$. } 
\begin{center}
\begin{tabular}{@{\extracolsep{0.2 cm}}ccccc}
\hline
\hline
Diquark & Mass & $E$ & Mass$^*$ & $E^*$ \\
\rule{0ex}{2.5ex}
$\{\tu\td\}_{1^-}$&1.44&0.28&1.33&0.50\\
\rule{0ex}{2.5ex}
$\{\tu\ts\}_{1^-}$&1.54&0.28&1.43&0.50\\
\rule{0ex}{2.5ex}
$\{\ts\ts\}_{1^-}$&1.64&0.27&1.54&0.50\\
\rule{0ex}{2.5ex}
$\{\tc\tu\}_{1^-}$&2.45&0.17&2.38&0.31\\
\rule{0ex}{2.5ex}
$\{\tc\ts\}_{1^-}$&2.54&0.16&2.48&0.30\\
\rule{0ex}{2.5ex}
$\{\tu\tb\}_{1^-}$&5.59&0.09&5.53&0.17\\
\rule{0ex}{2.5ex}
$\{\ts\tb\}_{1^-}$&5.67&0.09&5.62&0.16\\
\rule{0ex}{2.5ex}
$\{\tc\tb\}_{1^-}$&6.50&0.04&6.47&0.07\\
\rule{0ex}{2.5ex}
$\{\tc\tc\}_{1^-}$&3.42&0.07&3.38&0.13\\
\rule{0ex}{2.5ex}
$\{\tb\tb\}_{1^-}$&9.53&0.02&9.51&0.04\\
\hline
\hline
\end{tabular}
\end{center}
\end{table}
\section{Negative Parity Baryons}
\begin{figure}[htpb]
\vspace{-2cm}
       \centerline{\includegraphics[scale=0.38,angle=-90]{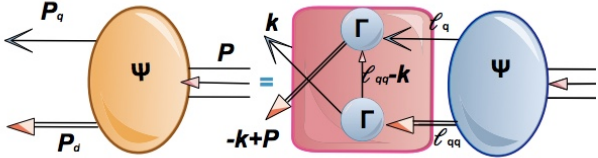}}
       \vspace{-2cm}\caption{Poincar\'e covariant FE to calculate baryon masses. The square represents the quark-diquark interaction Kernel. The single line denotes the dressed quark propagator, the double line is the diquark propagator while $\Gamma$ and $\Psi$ are the BSA and Faddeev amplitudes, respectively. Configuration of momenta is: $\ell_{qq}=-\ell + P$, $k_{qq}=-k+P$, $P=P_d+P_q$.}
       \label{faddevv-Fig1}
\end{figure}

In this section we extend the CI model to the heavy baryon sector of negative parity. We compute the masses of negative parity spin-$1/2$ and $3/2$ baryons composed of $\tu$, $\td$, $\ts$, $\tc$ and $\tb$ quarks in a quark-diaquark picture. We base our description of baryon bound-states on FE, which
is illustrated in Fig.~\ref{faddevv-Fig1}.
\vspace{1.0 cm}
\subsection{Baryons with Spin $1/2$}
The nucleon’s parity partner is composed
of pseudoscalar, vector, scalar and axial-vector diquark correlations and its Faddeev amplitude must change sign under
a parity transformation. These alterations lead to changes in the locations of the $\gamma_5$ matrices in the FE.
The Faddeev amplitude of the nucleon parity partner with the interaction employed in this article can be written in terms of:
\begin{align*}
  S(P)&=s(P)\gamma_5 I_D, & A_{\mu}^{i}(P)&=ia^{i}_{1}(P)\gamma_{\mu}-a_2^{i}(P)\hat{P}_\mu, \\
  S_{\mathpzc {p}}(P)&=-i{\mathpzc {p}}(P) I_D,&
  {\mathcal{V}}^{i}_{\mu}(P)&=iv^{i}_1(P)\gamma_5\gamma_\mu+v^{i}_2(P)\gamma_5 \hat{P}_{\mu}\;,\\
  \end{align*}
where $i=0,+$ and $\hat P^2=-1$. The mass of the ground-state baryon  with spin $1/2$ and negative parity comprised by the quarks $[\tq\tq\tqu]$ is  determined by a $10\times 10$ matrix FE.
In the explicit matrix representation, one can write it as follows~:
\begin{eqnarray}
\nonumber
\lefteqn{
 \left[ \begin{array}{r}
{\cal S}(P)\, u(P)\\
[0.7ex]
{\cal A}^i_\mu(P)\, u(P)\\
[0.7ex]
S_{\mathpzc {p}}(P)\, u(P)\\
[0.7ex]
 {\mathcal{V}}^{i}_{\mu}(P)\, u(P)
\end{array}\right]}\\ \nn \\
& =&  -\,4\,\int\frac{d^4\ell}{(2\pi)^4}\,{\cal M}(\ell;P)
\left[
\begin{array}{r}
{\cal S}(P)\, u(P)\\
[0.7ex]
{\cal A}^j_\nu(P)\, u(P)\\
[0.7ex]
S_{\mathpzc {p}}(P)\, u(P)\\

 {\mathcal{V}}^{i}_{\nu}(P)\, u(P)
\end{array}\right] ,\rule{1em}{0ex}
\label{FEone}
\end{eqnarray}
where $u(P)$ is a Dirac spinor; The kernel in Eq.~{\ref{FEone}} is
detailed in the appendix \ref{app:Fad}.
The general matrices ${\cal S}(P)$, ${\cal A}^i_\nu(P)$, $S_{\mathpzc {p}}(P)$ and ${\mathcal{V}}^{i}_{\nu}(P)$ which describe the momentum-space correlation between the quark and diquark in the nucleon and the Roper, are described in Refs.~\cite{Oettel:1998bk, Cloet:2007pi}.
The Faddeev amplitude is thus represented by the eigenvector~:
\bea
\Psi(P) =
(s ,
a_1^0,
a_1^+,
a_2^0,
a_2^+,
{\mathpzc {p}},
v^{0}_1,
v^{+}_1,
v^{0}_2,
v^{+}_2)^T .
\eea
We use {\em static approximation} for the exchanged quark with flavor $f$. It was introduced for the first time a long while ago in Ref.~\cite{Buck:1992wz}
\bea
S(p)=\frac{1}{i \gamma \cdot p + M_f} \to
\frac{1}{M_f} \,. \eea
A variation of it was implemented in~\cite{Xu:2015kta}, representing the quark propagator as
\bea
S(p)=\frac{1}{i \gamma \cdot p + M_f}\to
\frac{g^2_{N\,\Delta}}{i \gamma \cdot p + M_f}  \,. \eea
We follow refs.~\cite{Roberts:2011cf,Lu:2017cln,Chen:2012qr} and represent the quark (propagator) exchanged between the diquarks as a simple modification of the static approximation
\begin{equation}
S^{\rm T}(k) \to \frac{g_B}{M_f}\,.
\label{staticexchangec}
\end{equation}
The superscript ``T'' indicates matrix transpose.
In the implementation of this treatment for heavy baryons with spin-$1/2$ we use $g_B=1$. Explicit expressions for the flavor matrices $t$ for the diquark pieces can be found in Appendix~\ref{app:Fla}. The spin-$1/2$ heavy baryons are represented by the following column matrices:
 \begin{equation}
\nn \begin{array}{cc}
u_{\Xi_{cc}^{++} (\tu\tc\tc)}= \left[
\begin{array}{c}
[\tu\tc]\nop \tc \\
\{\tc\tc\}\mpo\tu\\
\{\tu\tc\}\mpo\tc\\
\left[\tu\tc\right]\noo\tc \\
\{\tc\tc\}\mni\tu\\
\{\tu\tc\}\mni\tc\\
\end{array} \right], &
u_{\Omega_{\tc\tc}^+(\ts\tc\tc)}=
\left[ \begin{array}{c}
[\ts\tc]\nop \tc \\
\{\tc\tc\}\mpo \ts \\
\{\ts\tc\}\mpo \tc \\
\left[\ts\tc\right]\noo \tc \\
\{\tc\tc\}\mni \ts \\
\{\ts\tc\}\mni \tc \\
\end{array} \right],
\end{array}
\end{equation}
\vspace{-0.5cm}
 \begin{equation}
\nn \begin{array}{cc}
u_{\Omega_{\tc}^0(\ts\ts\tc)}= \left[
\begin{array}{c}
[\ts\tc]\nop \ts \\
\{\ts\ts\}\mpo\tc\\
\{\ts\tc\}\mpo\ts\\
\left[\ts\tc\right]\noo \ts \\
\{\ts\ts\}\mni\tc\\
\{\ts\tc\}\mni\ts\\
\end{array} \right], &
u_{\Sigma_{\tc}^{++}(\tu\tu\tc)}=
\left[ \begin{array}{c}
[\tu\tc]\nop \tu \\
\{\tu\tu\}\mpo \tc \\
\{\tu\tc\}\mpo \tu \\
\left[\tu\tc\right]\noo \tu \\
\{\tu\tu\}\mni \tc \\
\{\tu\tc\}\mni \tu \\
\end{array} \right],
\end{array}
\end{equation}
\vspace{-0.5cm}
 \begin{equation}
\nn \begin{array}{cc}
u_{\Xi_{bb}^0(\tu\tb\tb)}= \left[
\begin{array}{c}
[\tu\tb]\nop \tb \\
\{\tb\tb\}\mpo\tu\\
\{\tu\tb\}\mpo\tb\\
\left[\tu\tb\right]\noo \tb \\
\{\tb\tb\}\mni\tu\\
\{\tu\tb\}\mni\tb\\
\end{array} \right], &
u_{\Omega_{bb}^-(\ts\tb\tb)}=
\left[ \begin{array}{c}
[\ts\tb]\nop \tb \\
\{\tb\tb\}\mpo \ts \\
\{\ts\tb\}\mpo \tb \\
\left[\ts\tb\right]\noo \tb \\
\{\tb\tb\}\mni \ts \\
\{\ts\tb\}\mni \tb \\
\end{array} \right],
\end{array}
\end{equation}
\vspace{-0.5cm}
 \begin{equation}
\nn \begin{array}{cc}
u_{\Omega_b^{-}(\ts\ts\tb)}= \left[
\begin{array}{c}
[\ts\tb]\nop \ts \\
\{\ts\ts\}\mpo\tb\\
\{\ts\tb\}\mpo\ts\\
\left[\ts\tb\right]\noo \ts \\
\{\ts\ts\}\mni\tb\\
\{\ts\tb\}\mni\ts\\
\end{array} \right], &
u_{\Sigma_b^+(\tu\tu\tb)}=
\left[ \begin{array}{c}
[\tu\tb]\nop \tu \\
\{\tu\tu\}\mpo \tb \\
\{\tu\tb\}\mpo \tu \\
\left[\tu\tb\right]\noo \tu \\
\{\tu\tu\}\mni \tb \\
\{\tu\tb\}\mni \tu \\
\end{array} \right],
\end{array}
\end{equation}
\vspace{-0.5cm}
 \begin{equation}
\nn \begin{array}{cc}
u_{\Omega(\tc\tc\tb)}= \left[
\begin{array}{c}
[\tc\tb]\nop \tc \\
\{\tc\tc\}\mpo\tb\\
\{\tc\tb\}\mpo\tc\\
\left[\tc\tb\right]\noo \tc \\
\{\tc\tc\}\mni\tb\\
\{\tc\tb\}\mni\tc\\
\end{array} \right], &
u_{\Omega(\tc\tb\tb)}=
\left[ \begin{array}{c}
[\tc\tb]\nop \tb \\
\{\tb\tb\}\mpo \tc \\
\{\tc\tb\}\mpo \tb \\
\left[\tc\tb\right]\noo \tb \\
\{\tb\tb\}\mni \tc \\
\{\tc\tb\}\mni \tb \\
\end{array} \right].
\end{array}
\end{equation}
Experimental and calculated masses of spin $1/2$-baryons with
charm and bottom quarks are listed in
Table~\ref{table-baryons-half}.
Our results for light baryons give masses larger than the expected values since our calculations for these states do not contain contributions associated with the meson cloud effect \cite{Roberts:1988yz} which work to reduce baryon masses.
The size of such corrections has been estimated:
for the nucleon, the reduction is roughly $0.2$ GeV and
for the $\Delta$ it is $0.16$ GeV. Our deliberately inflated masses allow us to achieve the correct results after incorporating the meson cloud effects.
If our calculations included these effects, our results would be modified to $m_N=0.98$ GeV, $m_{\Sigma}=1.20$ GeV and $m_{\Xi}=1.27$ GeV. This can effectively be achieved if we change our set of parameters as suggested in~\cite{Yin:2021uom}.
This is what we precisely do
for the case of heavy and heavy-light baryons $1/2^-$. These are less than 9\% different from the predicted value in~\cite{Qin:2019hgk} which already include the effects of the meson cloud. The values in column 4 of Table \ref{table-baryons-half} may vary slightly with the change of $g_B$.

\begin{table}[htbp]
\extrarowheight = -0.5ex
    \renewcommand{\arraystretch}{1.75}
\caption{\label{table-baryons-half} Spin 1/2 baryons. The results abbreviated by QRS are taken from Ref.~\cite{Qin:2019hgk}. In the fifth column the upper and lower limits indicate a change in mass due to a variation of $g_B=1\pm 0.5$ for heavy baryons.
The mass is greater when $g_B$ is smaller and decreases as $g_B$ increases.
In the case of light baryons we use $g_B$=1.18. For the case of baryons with negative parity we have only considered the contribution of dominant diquarks with the same parity. } 
\vspace{0 cm}
\begin{tabular}{@{\extracolsep{0.0 cm}}cccccc}
\hline
\hline
Baryon & (Exp.,+) & (CI,+)  & (Exp.,-) & (CI,-)& (QRS,-) \\
\rule{0ex}{3.5ex}
$N(\tu\tu\td)$ &0.94&1.14&1.54&1.82& 1.542\\
\rule{0ex}{3.5ex}
$\Sigma(\tu\tu\ts)$&1.19&1.36&1.75&1.96&1.581\\
\rule{0ex}{3.5ex}
$\Xi(\ts\tu\ts)$ &1.31&1.43&--&2.04&1.620\\
 \rule{0ex}{3.5ex}
$\Xi_{cc}^{++} (\tu\tc\tc)$&3.62&3.64&--& $3.80^{+0.3}_{-0.4}$ &3.790\\
\rule{0ex}{3.5ex}
$\Omega_{\tc\tc}^+(\ts\tc\tc)$ &--&3.76&& $3.95^{+0.3}_{-0.4}$ &3.829\\
\rule{0ex}{3.5ex}
$\Omega_{\tc}^0(\ts\ts\tc)$ &2.69&2.82&& $2.99^{+0.4}_{-0.5}$ &2.744\\
\rule{0ex}{3.5ex}
$\Sigma_{\tc}^{++}(\tu\tu\tc)$&2.45&2.58&& $2.64^{+0.2}_{-0.3}$ &2.666\\
\rule{0ex}{3.5ex}
$\Xi_{bb}^0(\tu\tb\tb)$ &--&10.06&& $10.17^{+0.3}_{-0.4}$ &10.289\\
\rule{0ex}{3.5ex}
$\Omega_{bb}^-(\ts\tb\tb)$ &--&10.14&& $10.32^{+0.3}_{-0.5}$ &10.328\\
\rule{0ex}{3.5ex}
$\Omega_b^{-}(\ts\ts\tb)$ &6.04&6.01&& $6.47^{+0.3}_{-0.3}$ &5.994\\
\rule{0ex}{3.5ex}
$\Sigma_b^+(\tu\tu\tb)$&5.81&5.78&& $6.36^{+0.3}_{-0.4}$ &5.916\\
\rule{0ex}{3.5ex}
$\Omega(\tc\tb\tb)$ &--&11.09&& $11.22^{+0.4}_{-0.5}$ &11.413\\
\rule{0ex}{3.5ex}
$\Omega(\tc\tc\tb)$ &--&8.01&& $8.17^{+0.2}_{-0.3}$&8.164\\
\hline
\hline
\end{tabular}
\end{table}
The masses of spin 1/2 baryons with only one heavy quark obey an equal spacing rule~\cite{GellMann:1962xb,Okubo:1961jc,Ebert:2005xj}:
\bea\label{eqGMO12}
    m_{\Sigma_q} + m_{\Omega_{q}} = 2m_{\Xi_q}. \quad q=\tc,\ \tb.
\eea
With this equation we can predict the mass of the baryons $ \Xi (\tu\ts\tc)$ and $ \Xi(\tu\ts\tb)$ in  Table \ref{tabla:SigmaOmegaXi12}.
\begin{table}[H]
  \centering
  \caption{Masses of spin $1/2$ baryons predicted by CI. The results abbreviated by QRS are taken from Ref.~\cite{Qin:2019hgk}}.
            \begin{tabular}{@{\extracolsep{0.3cm}}c|c|c|c|c}
            \hline \hline
             &(CI,+) &  (Exp., +)& (CI, -) & (QRS,-)  \\ \hline
           $m_{\Xi^{+}_{\tc}(\tu\ts\tc)}$& 2.70 & 2.47& 2.81 & 2.70 \\
           \rule{0ex}{2.5ex}
           $m_{\Xi^{0}_{\tb}(\tu\ts\tb)}$ & 5.89 & 5.80 & 6.42&5.96\\
             \hline \hline
            \end{tabular}
            \label{tabla:SigmaOmegaXi12}
 \end{table}
 %
\vspace{-1cm}
\subsection{Baryons with Spin $3/2$}
Baryons with spin $3/2$ are specially important because they can involve states with three $\tc$-quarks and three $\tb$-quarks. In order to calculate the masses we note that it is not possible to combine a spin-zero
diquark with a spin-1/2 quark to obtain spin-$3/2$ baryon. Hence such a baryon is comprised solely of vector correlations.
Understanding the structure of these states is thus simpler in some sense than the nucleon.
The Faddeev amplitude for the positive-energy baryon
is~:
\bea
\Psi_\mu = \psi_{\mu\nu}(P)u_\nu(P) \;, \nn
\eea
where $P$ is the baryon's total momentum and $u_\nu(P)$ is a Rarita-Schwinger spinor,
\bea \label{fd1}
\psi_{\mu\nu}(P)u_{\nu} &=& \Gamma_{qq_{1^+} \mu} \Delta_{\mu\nu,_{qq}}^{1^+}(\ell_{qq}){\cal D}_{\nu\rho}(P)u_{\rho}(P) \,
\eea
and
\bea
\label{DeltaFA}
{\cal D}_{\nu\rho}(\ell;P) &=& {\cal S}(\ell;P) \, \delta_{\nu\rho} + \gamma_5{\cal A}_\nu(\ell;P) \,\ell^\perp_\rho \,.
\eea
We give more details of this equation in the  appendix~\ref{App:EM}.
%
We assume that the parity partner of a given baryon is obtained by replacing the diquark correlation(s) involved by its (their) parity partners.
We consider the baryons with two possible structures: $\tq\tq\tq$ and $\tqu\tq\tq$. \\ \\
%
\noindent
{\bf Baryons($\tq\tq\tq$):} There exists only a single possible combination of diquarks for a baryon composed of the same three quarks $(\tq\tq\tq)$. The Faddeev amplitude for this case is:
\begin{equation}
{\cal D}_{\nu\rho}(\ell;P) u_\rho^B(P) = f^B(P) \, \gc\, u_\nu^B(P)\,.
\label{DnurhoI}
\end{equation}
Employing Feynman rules for Fig.~\ref{faddevv-Fig1} and using the expression for the Faddev amplitude, Eq.~({\ref{DnurhoI}}), we can write
\bea
 f^B(P)\gc u_\mu^B(P)=
4\frac{g_B}{M_{\tq}}\int \frac{d^4 \ell}{(2\pi)^4}\mathcal{M}f^B(P)\gc u_\nu^B(P) \,,
\eea
where we have suppressed the functional dependence of $\mathcal{M}$ on
momenta for the simplicity of notation. We
now multiply both sides by ${\bar{u}}^B_{\beta}(P)\gc$ from the left and sum over the polarization,
not explicitly shown here, to obtain
\bea\nn
\gc\Lambda_{+}(P)R_{\mu\beta}(P)\gc=4\frac{g_B}{M_{\tq}}\int \frac{d^4 \ell}{(2\pi)^4}\mathcal{M}\gc\Lambda_{+}(P)R_{\mu\beta}\gc.\\
\eea
Finally we contract with $\delta_{\mu\beta}$
\begin{eqnarray} \nn  \label{o-diquark}
2\pi^2 = \frac{1}{M_{\tq}}\frac{E_{\{\tq\tq\}_{1^-}}^2}{m_{\{\tq\tq\}_{1^-}}^2}
 \hspace{-2mm} \int_0^1 \hspace{-2mm} d\alpha\, {\cal L}^B \overline{\cal C}^{\rm iu}_1(\omega (\alpha,M_{\tq}^2,m^2_{\{\tq\tq\}_{1^-}},m_B^2))\,,
\end{eqnarray}
where we have defined
\bea
{\cal L}^B= [m_{\{\tq\tq\}_{1^-}}^2 + (1-\alpha)^2 m_B^2][\alpha m_B \mn M_{\tq}]
\,.
\eea
From the last two expressions, it is straightforward to compute the mass of the baryon constituted
by three equally heavy quarks. \\ \\
\noindent
{\bf Baryons($\tqu\tq\tq$):}
 For a baryon with quark structure ($\tqu\tq\tq$), there are two possible diquarks, $\{\tq\tq\}_{1^-}$ and $\{\tqu\tq\}_{1^-}$. The Faddeev amplitude for such a baryon is:
\begin{equation}  \label{d-diquark}
{\cal D}_{\nu\mu}^{B}(P)u_{\mu}^{B}(P;s) =
\sum_{i} d^{i}(P) \delta_{\nu\lambda}\gc u_{\lambda}^{B}(P;s),
\end{equation}
where $i=\{\tqu\tq\}_{1^-},\{\tq\tq\}_{1^-}$. The FE has the form
\begin{equation}
\left[\begin{matrix}
d^{\{\tqu\tq\}_{1^-}}\\ d^{\{\tq\tq\}_{1^-}}\end{matrix}\right]\gc
u_{\mu}^{B}= -4 \int \frac{d^{4}l}{(2\pi)^{4}} \\
{\cal M}
\left[\begin{matrix}
d^{\{\tqu\tq\}_{1^-}}\\ d^{\{\tq\tq\}_{1^-}}\end{matrix}\right]\gc u_{\nu}^{B} \,,
\label{eq:D2}
\end{equation}%
where
\bea
\cal{M}=\left[\begin{matrix}
{\cal M}_{\mu\nu}^{\{\tqu\tq\}_{1^-},\{\tqu \tq\}_{1^-}} & {\cal M}_{\mu\nu}^{\{\tqu \tq\}_{1^-},\{\tq\tq\}_{1^-}} \\
{\cal M}_{\mu\nu}^{\{\tq\tq\}_{1^-},\{\tqu \tq\}_{1^-}} & {\cal M}_{\mu\nu}^{\{\tq\tq\}_{1^-},\{\tq\tq\}_{1^-}}
\end{matrix}\right]
\eea
with the elements of the matrix ${\cal M}$ given by~:
\begin{equation} \nn
\begin{split}
{\cal M}_{\mu\nu}^{00}& =t^{f_{00}}\frac{1}{M_{\tqu}} \,
\Gamma_{\rho}^{1^{-}}(\ell_{\tqu\tq})  \,
\bar{\Gamma}_{\mu}^{1^{-}}(-k_{\tqu\tq}) \, S(l_{\tq}) \,
\Delta_{\rho\nu}^{1^{+}}(\ell_{\tqu\tq}), \\
{\cal M}_{\mu\nu}^{01}&=t^{f_{01}}\frac{1}{M_{\tq}} \,
\Gamma_{\rho}^{1^{-}}(\ell_{\tq\tq})  \,
\bar{\Gamma}_{\mu}^{1^{-}}(-k_{\tqu\tq}) \, S(l_{q_1}) \,
\Delta_{\rho\nu,\{\tq\tq\}}^{1^{+}}(\ell_{\tq\tq}),
\end{split}
\end{equation}
\begin{equation}
\begin{split}
{\cal M}_{\mu\nu}^{10} &=t^{f_{10}}\frac{1}{M_{\tq}} \,
\Gamma_{\rho}^{1^{-}}(\ell_{\tqu\tq})  \,
\bar{\Gamma}_{\mu}^{1^{-}}(-k_{\tq\tq}) \, S(l_{\tq}) \,
\Delta_{\rho\nu}^{1^{+}}(\ell_{\tqu\tq}), \\
{\cal M}_{\mu\nu}^{11}&=t^{f_{11}} \frac{1}{M_{\tq}} \,
\Gamma_{\rho}^{1^{-}}(\ell_{\tq\tq})  \,
\bar{\Gamma}_{\mu}^{1^{-}}(-k_{\tq\tq}) \, S(l_{\tq_1}) \,
\Delta_{\rho\nu}^{1^{+}}(\ell_{\tq\tq}),
\end{split}
\end{equation}
where $t^f$ are the flavor matrices and can be found in appendix~\ref{app:Fla}. The color-singlet bound states constructed from three heavy charm/bottom quarks are:
\begin{eqnarray}
&&\nn \begin{array}{cc}
u_{\Omega_{\tc\tc\tc}^{++*}}= \left[
\begin{array}{c}
\{\tc\tc\}\mni \tc \\
\end{array} \right], &
\hspace{1.0cm}u_{\Omega_{\tb\tb\tb}^{-*}}=
\left[ \begin{array}{c}
\{\tb\tb\}\mni \tb \\
\end{array} \right],
\end{array} \\ \nn \\
&&\nn \begin{array}{cc}
u_{\Omega_{\tc\tc\tb}^{+*}}= \left[
\begin{array}{c}
\{\tc\tc\}\mni\tb \\
\{\tc\tb\}\mni \tc \\
\end{array} \right], &
\hspace{1cm}u_{\Omega_{\tc\tb\tb}^{0*}}=
\left[ \begin{array}{c}
\{\tc\tb\}\mni \tb \\
\{\tb\tb\}\mni \tc
\end{array} \right].
\end{array} \\
\end{eqnarray}
The column vectors representing singly and doubly heavy baryons are:
\begin{eqnarray}
&&\nn \hspace{-0.5cm} \begin{array}{cc}
u_{\Sigma_{\tc}^{++*}{(\tu\tu\tc)}}=
\left[ \begin{array}{c}
\{\tu\tu\}\mni \tc \\
\{\tu\tc\}\mni \tu
\end{array} \right],
&
\hspace{0.3cm}u_{\Xi_{\tc\tc}^{++*}{(\tu\tc\tc)}}= \left[
\begin{array}{c}
\{\tu\tc\}\mni \tc \\
\{\tc\tc\}\mni \tu \\
\end{array} \right],
\end{array} \\ \nn \\
&&\nn \hspace{-0.2cm} \begin{array}{cc}
u_{\Omega_{\tc}^{0*}{(\ts\ts\tc)}}=
\left[ \begin{array}{c}
\{\ts\ts\}\mni \tc \\
\{\ts\tc\}\mni \ts
\end{array} \right],
 &
\hspace{0.6cm}u_{\Omega_{\tc\tc}^{+*}{(\ts\tc\tc)}}= \left[
\begin{array}{c}
\{\ts\tc\}\mni \tc \\
\{\tc\tc\}\mni \ts \\
\end{array} \right],
\end{array}
\\ \nn \\
&&\nn \hspace{-0.3cm} \begin{array}{cc}
u_{\Sigma_{\tb}^{+*}{(\tu\tu\tb)}}=
\left[ \begin{array}{c}
\{\tu\tu\}\mni \tb \\
\{\tu\tb\}\mni \tu
\end{array} \right],
&
\hspace{0.5cm}u_{\Sigma_{\tb\tb}^{0*}{(\tu\tb\tb)}}= \left[
\begin{array}{c}
\{\tu\tb\}\mni\tb \\
\{\tb\tb\}\mni \tu \\
\end{array} \right],
\end{array} \\ \nn \\
&&\nn \hspace{-0.2cm} \begin{array}{cc}
u_{\Omega_{\tb}^{-*}{(\ts\ts\tb)}}=
\left[ \begin{array}{c}
\{\ts\ts\}\mni \tb \\
\{\ts\tb\}\mni \ts
\end{array} \right],
&
\hspace{0.5cm}u_{\Omega_{\tb\tb}^{-*}{(\ts\tb\tb)}}= \left[
\begin{array}{c}
\{\ts\tb\}\mni\tb \\
\{\tb\tb\}\mni \ts \\
\end{array} \right].
\end{array}
\end{eqnarray}
We have solved FE~(\ref{fd1}) and
obtained masses and eigenvectors of the ground-state baryons of spin $3/2$ using $g_B=1$. The results are listed in
Table~\ref{table-baryons-oct} below:

\begin{table}[htbp]
\extrarowheight = -0.5ex
    \renewcommand{\arraystretch}{1.75}
\caption{\label{table-baryons-oct} Masses of baryons with spin 3/2 in GeV. The results denoted by QRS have been taken from~\cite{Qin:2019hgk}. Experimental results have been labelled with an asterisk. The last table is listed in units of $m_{\Omega_{\tc\tc\tc}}$} 
\vspace{0 cm}
\begin{tabular}{@{\extracolsep{0.0 cm}}cccccc}
\hline
\hline
Baryon & (Lat.,+) & (CI,+)  & (Exp.,-) & (CI,-)& (QRS,-) \\
$\Delta(\tu\tu\tu)$&1.23*&1.39&1.65&2.07&1.726\\
\rule{0ex}{3.5ex}
$\Sigma^*(\tu\tu\ts)$&1.39*&1.51&1.67&2.16&1.785\\
\rule{0ex}{3.5ex}
$\Xi^*(\ts\tu\ts)$&1.53*&1.63&1.82&2.26&1.843\\
\rule{0ex}{3.5ex}
$\Omega(\ts\ts\ts)$&1.67*&1.76&--&2.36&1.902\\
\hline
\hline
 $\Omega_{\tc\tc\tc}^{++*}$  &4.80&4.93&& $5.28^{+0.04}_{-0.02}$&5.027\\
   \rule{0ex}{3.5ex}
 $\Omega_{\tb\tb\tb}^{-*}$  &14.37&14.23&& $14.39^{+0.1}_{-0.02}$&14.771\\
  \rule{0ex}{3.5ex}
 $\Omega_{\tc\tc\tb}^{+*}$  &8.01&8.03&& $8.28^{+0.03}_{-0.01}$ &8.275\\
   \rule{0ex}{3.5ex}
 $\Omega_{\tc\tb\tb}^{0*}$  &11.20&11.12&& $11.35^{+0.01}_{-0.01}$ &11.523\\
 \hline
  \hline
 $\Sigma_{\tc}^{++*}{(\tu\tu\tc)}$ &0.53*&0.57&& $0.67^{+0.01}_{-0.01}$ &0.59\\
 \rule{0ex}{3.5ex}
 $\Xi_{\tc\tc}^{++*}{(\tu\tc\tc)}$  &0.75&0.79&& $0.89^{+0.1}_{-0.1}$ &0.83\\
 \rule{0ex}{3.5ex}
 $\Omega_{\tc}^{0*}{(\ts\ts\tc)}$  &0.58*&0.61&& $0.72^{+0.02}_{-0.02}$ &0.63\\
 \rule{0ex}{3.5ex}
 $\Omega_{\tc\tc}^{+*}{(\ts\tc\tc)}$ &0.78&0.82&& $0.92^{+0.04}_{-0.02}$ &0.84\\
  \rule{0ex}{3.5ex}
 $\Sigma_{\tb}^{+*}{(\tu\tu\tb)}$ &1.21*&1.23&& $1.32^{+0.02}_{-0.03}$ &1.28\\
  \rule{0ex}{3.5ex}
 $\Xi_{bb}^{0*}{(\tu\tb\tb)}$ &2.11&2.12&& $2.10^{+0.02}_{-0.02}$ &2.19\\
  \rule{0ex}{3.5ex}
 $\Omega_{\tb}^{-*}{(\ts\ts\tb)}$&1.26&1.28&& $1.52^{+0.03}_{-0.02}$ &1.30\\
  \rule{0ex}{3.5ex}
 $\Omega_{\tb\tb}^{-*}{(\ts\tb\tb)}$  &2.14&2.10&& $2.10^{+0.02}_{-0.02}$ &2.20\\
\hline
\hline
\end{tabular}
\end{table}
For baryons of spin 3/2 with three distinct flavors of quarks we can again use equation (\ref {eqGMO12}). The results are depicted in Table~\ref{tabla:SigmaOmegaXi12}. As in the case of baryons with spin $1/2$, the masses of light baryons with spin $3/2$ are deliberately inflated to leave room for the
contributions of the meson cloud.
\begin{table}[H]
  \centering
  \caption{Mass Predicted by our model for baryons with spin $3/2$. The results abbreviated by QRS are taken from Ref.~\cite{Qin:2019hgk}}.
            \begin{tabular}{@{\extracolsep{0.3 cm}}c|c|c|c|c}
            \hline \hline
             &(CI,+) &  (Exp., +)& (CI, -) & (QRS,-)  \\ \hline
           $m_{\Xi^{+}_{\tc}(\tu\ts\tc)}$& 2.83 & 2.65& 3.34 & 2.93 \\
           \rule{0ex}{2.5ex}
           $m_{\Xi^{0}_{\tb}(\tu\ts\tb)}$& 6.02 & 5.95 & 6.82&6.20\\
             \hline \hline
            \end{tabular}
            \label{tabla:SigmaOmegaXi12}
 \end{table}
  Our percentage difference with the values obtained in \cite{Qin:2019hgk} for baryons with $3/2^-$
 are less than 6\% for the majority.

\section{Conclusions}
\label{Conclusions}

The CI model was first introduced in~\cite{GutierrezGuerrero:2010md}. It adapts itself well to the infrared
behavior of QCD. It incorporates a mass scale of about 500 MeV for the gluon, mimics confinement through
the absence of quark production thresholds, respects the axial vector Ward-Takahashi identity and preserves low energy Golberger-Treiman relations. Therefore, it is able to reproduce hadron spectrum and masses to a desirable accuracy. Most of the meson and baryon masses containing light and heavy quarks have already been reported in literature using the CI. In this article, we compute the masses of the remaining negative parity baryons containing heavy quarks. The set of parameters we adopt is inspired by a quark-mass fit studied in our previous work~\cite{Raya:2017ggu}. In the quark-diquark picture of baryons, we need to evaluate several meson and diquark masses through the BSE before embarking upon the
evaluation of baryon masses through the FE. The Tables~\ref{parameters},~\ref{table-M} show the parameters used in the entirety of
the article. Tables~\ref{table-mesones-pseudo},~\ref{table-mesones-esc},~\ref{table-mesones-vec},~\ref{table-mesones-av} depict all the meson masses. The diquark masses are detailed in Tables~\ref{table-diquarks-scalar},~\ref{table-di-pscalar},~\ref{table-diquarks-Av},~\ref{table-Vector-diquaks}. For bound states of two particles, we have calculated the masses of about 80 particles.\\
The spin-1/2 and spin-3/2 heavy baryon masses are listed in Table~\ref{table-baryons-half} and Table~\ref{table-baryons-oct}, respectively. Motivated by our earlier satisfactory computation of the singly and doubly heavy positive parity baryons,
and our comparisons of the negative parity baryon masses in the present work with other established models whenever possible, we are confident
our predictions will lie in close proximity of the future experimental observations.
The computed masses in this article include 58 baryons, 29 with positive parity and 29 with negative parity. In total, we present the computation of approximately 138 states.
Our planned next steps of research will involve computation of excited states, tetra- and penta-quarks, as well as form factors of
mesons and baryons containing heavy quarks.
\vspace{-1cm}
\begin{acknowledgements}
 We acknowledge useful discussions with Marco
 Bedolla.
 L. X. Guti\'errez-Guerrero wishes to thank the National Council for Science and Technology of Mexico (CONACyT) for the support provided to her through the programme of C\'atedras CONACyT. This research was also partly supported by Coordinaci\'on de la Investigaci\'on Cientifica (CIC) of the University of Michoacan and CONACyT, Mexico, through Grant nos. 4.10 and CB2014-22117, respectively. \\
\end{acknowledgements}

\appendix
\setcounter{equation}{0}
\renewcommand{\theequation}{\Alph{section}.\arabic{equation}}
\vspace{-1cm}
\setcounter{equation}{0}
\section{Euclidean Space Conventions}
\label{App:EM}
In our Euclidean space formulation:
\begin{equation}
p\cdot q=\sum_{i=1}^4 p_i q_i\,;
\end{equation}
where
\begin{eqnarray}\nn
&&\{\gamma_\mu,\gamma_\nu\}=2\,\delta_{\mu\nu}\,;\;
\gamma_\mu^\dagger = \gamma_\mu\,;\;
\sigma_{\mu\nu}= \frac{i}{2}[\gamma_\mu,\gamma_\nu]\,; \; \\
&&{\rm tr}\,[\gamma_5\gamma_\mu\gamma_\nu\gamma_\rho\gamma_\sigma]=
-4\,\epsilon_{\mu\nu\rho\sigma}\,, \epsilon_{1234}= 1\,.
\end{eqnarray}
A positive energy spinor satisfies
\begin{equation}
\bar u(P,s)\, (i \gamma\cdot P + M) = 0 = (i\gamma\cdot P + M)\, u(P,s)\,,
\end{equation}
where $s=\pm$ is the spin label.  It is conventionally normalised as~:
\begin{equation}
\bar u(P,s) \, u(P,s) = 2 M \,,
\end{equation}
and may be expressed explicitly as~:
\begin{equation}
u(P,s) = \sqrt{M- i {\cal E}}
\left(
\begin{array}{l}
\hspace{0.5cm} \chi_s\\
\displaystyle \frac{\vec{\sigma}\cdot \vec{P}}{M - i {\cal E}} \chi_s
\end{array}
\right)\,,
\end{equation}
with ${\cal E} = i \sqrt{\vec{P}^2 + M^2}$,
\begin{equation}
\chi_+ = \left( \begin{array}{c} 1 \\ 0  \end{array}\right)\,,\;
\chi_- = \left( \begin{array}{c} 0\\ 1  \end{array}\right)\,.
\end{equation}
For the free-particle spinor, $\bar u(P,s)= u(P,s)^\dagger \gamma_4$.
It can be used to construct a positive energy projection operator:
\begin{equation}
\label{Lplus} \Lambda_+(P):= \frac{1}{2 M}\,\sum_{s=\pm} \, u(P,s) \, \bar
u(P,s) = \frac{1}{2M} \left( -i \gamma\cdot P + M\right).
\end{equation}
A negative energy spinor satisfies
\begin{equation}
\bar v(P,s)\,(i\gamma\cdot P - M) = 0 = (i\gamma\cdot P - M) \, v(P,s)\,,
\end{equation}
and possesses properties and satisfies constraints obtained through obvious analogy
with $u(P,s)$. A charge-conjugated BSA is obtained via
\begin{equation}
\label{chargec}
\bar\Gamma(k;P) = C^\dagger \, \Gamma(-k;P)^{\rm T}\,C\,,
\end{equation}
where ``T'' denotes transposing  all matrix indices and
$C=\gamma_2\gamma_4$ is the charge conjugation matrix, $C^\dagger=-C$.  Moreover, we note that
\begin{equation}
C^\dagger \gamma_\mu^{\rm T} \, C = - \gamma_\mu\,, \; [C,\gamma_5] = 0\,.
\end{equation}
We employ a Rarita-Schwinger spinor to represent a covariant spin-$3/2$ field.  The positive energy
spinor is defined by the following equations:
\begin{equation}
\label{rarita}
(i \gamma\cdot P + M)\, u_\mu(P;r) = 0\,,
\gamma_\mu u_\mu(P;r) = 0\,,
P_\mu u_\mu(P;r) = 0,
\end{equation}
where $r=-3/2,-1/2,1/2,3/2$.  It is normalised as:
\begin{equation}
\bar u_{\mu}(P;r^\prime) \, u_\mu(P;r) = 2 M\,,
\end{equation}
and satisfies a completeness relation
\begin{equation}
\label{Deltacomplete}
\frac{1}{2 M}\sum_{r=-3/2}^{3/2} u_\mu(P;r)\,\bar u_\nu(P;r) =
\Lambda_+(P)\,R_{\mu\nu}\,,
\end{equation}
where
\begin{equation}
R_{\mu\nu} = \delta_{\mu\nu} \mbox{\boldmath $I$}_{\rm D} -\frac{1}{3} \gamma_\mu \gamma_\nu +
\frac{2}{3} \hat P_\mu \hat P_\nu \mbox{\boldmath $I$}_{\rm D} - i\frac{1}{3} [ \hat P_\mu
\gamma_\nu - \hat P_\nu \gamma_\mu]\,,
\end{equation}
with $\hat P^2 = -1$. It is very useful in simplifying the FE for a positive energy decouplet state.
\section{Kernel in FE}\label{app:Fad}
 \begin{align*}
\nn &M^{11}= t^{\tq T}t^{[\tq\tqu]_{0^+}}t^{[\tq\tqu]_{0^+}T}t^{\tq}\\ \nn
&\times \{\textcolor{magenta}{\g^{0^+}_{[\tq\tqu]}(l_{\tq\tqu}) S_{\tqu}^T\overline{\g}^{0^+}_{[\tq\tqu]}(-k_{\tq\tqu})S_{\tq}(l_{\tq})\D^{0^+}_{[\tq\tqu]}(l_{\tq\tqu})}\}\\
\nn &M^{12}_\nu= t^{\tq T}t^{\{\tq\tq\}_{1^+}}t^{[\tq\tqu]_{0^+}T}t^{\tqu}\\
&\nn\times\{\textcolor{magenta}{\g^{1^+}_{\{\tq\tq\},\mu}(l_{\tq\tq}) S_{\tq}^T\overline{\g}^{0^+}_{[\tq\tqu]}(-k_{\tq\tqu})S_{\tqu}(l_{\tqu})\D^{1^+}_{\{\tq\tq\},\mu\nu}(l_{\tq\tq})}\}\\
&\nn M_\nu ^{13}= t^{\tq T}t^{\{\tq\tqu\}_{1^+}}t^{[\tq\tqu]_{0^+}T}t^{\tq}\\
&\nn\times\{\textcolor{magenta}{\g^{1^+}_{\{\tq\tqu\},\mu}(l_{\tq\tqu}) S_{\tqu}^T\overline{\g}^{0^+}_{[\tq\tqu]}(-k_{\tq\tqu})S_{\tq}(l_{\tq})\D^{1^+}_{\{\tq\tqu\},\mu\nu}(l_{\tq\tqu})}\}\\
\nn & M^{16}= t^{\tq T}t^{[\tq\tqu]_{0^-}}t^{[\tq\tqu]_{0^+}T}t^{\tq}\\
&\nn \times\{\textcolor{magenta}{\g^{0^-}_{[\tq\tqu]}(l_{\tq\tqu}) S_{\tqu}^T\overline{\g}^{0^+}_{[\tq\tqu]}(-k_{\tq\tqu})S_{\tq}(l_{\tq})\D^{0^-}_{[\tq\tqu]}(l_{\tq\tqu})}\}\\
\nn &M^{17}_\nu= t^{\tq T}t^{\{\tq\tq\}_{1^-}}t^{[\tq\tqu]_{0^+}T}t^{\tqu}\\
&\nn\times\{\textcolor{magenta}{\g^{1^-}_{\{\tq\tq\},\mu}(l_{\tq\tq}) S_{\tq}^T\overline{\g}^{0^+}_{[\tq\tqu]}(-k_{\tq\tqu})S_{\tqu}(l_{\tqu})\D^{1^-}_{\{\tq\tq\},\mu\nu}(l_{\tq\tq})}\}\\
&\nn M_\nu^{18} = t^{\tq T}t^{\{\tq\tqu\}_{1^-}}t^{[\tq\tqu]_{0^+}T}t^{\tq}\\
&\nn\times\{\textcolor{magenta}{\g^{1^-}_{\{\tq\tqu\},\mu}(l_{\tq\tqu}) S_{\tqu}^T\overline{\g}^{0^+}_{[\tq\tqu]}(-k_{\tq\tqu})S_{\tq}(l_{\tq})\D^{1^-}_{\{\tq\tqu\},\mu\nu}(l_{\tq\tqu})}\}\\
\nn &M_{\mu}^{21}=t^{\tqu T}t^{[\tq\tqu]_{0^+}}t^{\{\tq\tq\}_{1^+}T}t^{\tq}\\
\nn &\times\{\textcolor{magenta}{\{\g^{0^+}_{[\tq\tqu]}(l_{\tq\tqu}) S_{\tq}^T\overline{\g}^{1^+}_{\{\tq\tq\},\mu}(-k_{\tq\tq})S_{\tq}(l_{\tq})\D^{0^+}_{[\tq\tqu]}(l_{\tq\tqu})\}}\\
\nn & M_{\mu\nu}^{22}=t^{\tqu T}t^{\{\tq\tq\}_{1^+}}t^{\{\tq\tq\}_{1^+}T}t^{\tqu}\\
&\nn \times\{\textcolor{magenta}{\g^{1^+}_{\{\tq\tq\},\rho}(l_{\tq\tq}) S_{\tq}^T\overline{\g}^{1^+}_{\{\tq\tq\},\mu}(-k_{\tq\tq})S_{\tq}(l_{\tq})\D^{1^+}_{\{\tq\tq\},\rho\nu}(l_{\tq\tq})}\}\\
\nn & M_{\mu\nu}^{23}=t^{\tqu T}t^{\{\tq\tqu\}_{1^+}}t^{\{\tq\tq\}_{1^+}T}t^{\tq}\\
\nn &\times\{\textcolor{magenta}{\g^{1^+}_{\{\tq\tqu\},\rho}(l_{\tq\tqu}) S_{\tq}^T\overline{\g}^{1^+}_{\{\tq\tq\},\mu}(-k_{\tq\tq})S_{\tq}(l_{\tq})\D^{1^+}_{\{\tq\tqu\},\rho\nu}(l_{\tq\tqu})}\}\\
\nn & M_{\mu}^{26}=t^{\tqu T}t^{[\tq\tqu]_{0^-}}t^{\{\tq\tq\}_{1^+}T}t^{\tq}\\
\nn&\times\{\textcolor{magenta}{\g^{0^-}_{[\tq\tqu]}(l_{\tq\tqu}) S_{\tq}^T\overline{\g}^{1^+}_{\{\tq\tq\},\mu}(-k_{\tq\tq})S_{\tq}(l_{\tq})\D^{0^-}_{[\tq\tqu]}(l_{\tq\tqu})}\}\\
\nn & M_{\mu\nu}^{27}=t^{\tqu T}t^{\{\tq\tq\}_{1^-}}t^{\{\tq\tq\}_{1^+}T}t^{\tqu}\\
&\nn \times\{\textcolor{magenta}{\g^{1^-}_{\{\tq\tq\},\rho}(l_{\tq\tq}) S_{\tq}^T\overline{\g}^{1^+}_{\{\tq\tq\},\mu}(-k_{\tq\tq})S_{\tq}(l_{\tq})\D^{1^-}_{\{\tq\tq\},\rho\nu}(l_{\tq\tq})}\}\\
\end{align*}
\begin{align*}
\nn &M_{\mu\nu}^{28} =t^{\tqu T}t^{\{\tq\tqu\}_{1^-}}t^{\{\tq\tq\}_{1^+}T}t^{\tq}
\\ \nn &\times\{\textcolor{magenta}{\g^{1^-}_{\{\tq\tqu\},\rho}(l_{\tq\tqu}) S_{\tq}^T\overline{\g}^{1^+}_{\{\tq\tq\},\mu}(-k_{\tq\tq})S_{\tq}(l_{\tq})\D^{1^-}_{\{\tq\tqu\},\rho\nu}(l_{\tq\tqu})}\}\\
\nn & M_{\mu}^{31}=t^{\tq T}t^{[\tq\tqu]_{0^+}}t^{\{\tq\tqu\}_{1^+}T}t^{\tq}\\
\nn &\times\{\textcolor{magenta}{\g^{0^+}_{[\tq\tqu]}(l_{\tq\tqu}) S_{\tqu}^T\overline{\g}^{1^+}_{\{\tq\tqu\},\mu}(-k_{\tq\tqu})S_{\tq}(l_{\tq})\D^{0^+}_{[\tq\tqu]}(l_{\tq\tqu})}\}=\mathcal{K}^{31}_{\mu}\\
\nn& M_{\mu\nu}^{32}=t^{\tq T}t^{\{\tq\tq\}_{1^+}}t^{\{\tq\tqu\}_{1^+}T}t^{\tqu}\\
&\times\{\textcolor{magenta}{\g^{1^+}_{\{\tq\tq\},\rho}(l_{\tq\tq}) S_{\tq}^T\overline{\g}^{1^+}_{\{\tq\tqu\},\mu}(-k_{\tq\tqu})S_{\tqu}(l_{\tqu})\D^{1^+}_{\{\tq\tq\},\rho\nu}(l_{\tq\tq})}\}\\
\nn & M_{\mu\nu}^{33}=t^{\tq T}t^{\{\tq\tqu\}_{1^+}}t^{\{\tq\tqu\}_{1^+}T}t^{\tq}\\
\nn &\times\{\textcolor{magenta}{\g^{1^+}_{\{\tq\tqu\},\rho}(l_{\tq\tqu}) S_{\tq}^T\overline{\g}^{1^+}_{\{\tq\tqu\},\mu}(-k_{\tq\tqu})S_{\tq}(l_{\tq})\D^{1^+}_{\{\tq\tqu\},\rho\nu}(l_{\tq\tqu})}\}\\
\nn & M_{\mu}^{36}=t^{\tq T}t^{[\tq\tqu]_{0^-}}t^{\{\tq\tqu\}_{1^+}T}t^{\tq}\\
\nn &\times\{\textcolor{magenta}{\g^{0^-}_{[\tq\tqu]}(l_{\tq\tqu}) S_{\tq}^T\overline{\g}^{1^+}_{\{\tq\tqu\},\mu}(-k_{\tq\tqu})S_{\tqu}(l_{\tqu})\D^{0^-}_{[\tq\tqu]}(l_{\tq\tqu})}\}\\
\nn& M_{\mu\nu}^{37}=t^{\tq T}t^{\{\tq\tq\}_{1^-}}t^{\{\tq\tqu\}_{1^+}T}t^{\tqu}\\
&\times\{\textcolor{magenta}{\g^{1^-}_{\{\tq\tq\},\rho}(l_{\tq\tq}) S_{\tq}^T\overline{\g}^{1^+}_{\{\tq\tqu\},\mu}(-k_{\tq\tqu})S_{\tqu}(l_{\tqu})\D^{1^-}_{\{\tq\tq\},\rho\nu}(l_{\tq\tq})}\}\\
\nn & M_{\mu\nu}^{38}=t^{\tq T}t^{\{\tq\tqu\}_{1^-}}t^{\{\tq\tqu\}_{1^+}T}t^{\tq}\\
\nn & \times\{\textcolor{magenta}{\g^{1^-}_{\{\tq\tqu\},\rho}(l_{\tq\tqu}) S_{\tq}^T\overline{\g}^{1^+}_{\{\tq\tqu\},\mu}(-k_{\tq\tqu})S_{\tq}(l_{\tq})\D^{1^-}_{\{\tq\tqu\},\rho\nu}(l_{\tq\tqu})}\}\\
\nn& M^{61}=t^{\tq T}t^{[\tq\tqu]_{0^+}}t^{[\tq\tqu]_{0^-}T}t^{\tq}\\
\nn &\times\{\textcolor{magenta}{\{\g^{0^+}_{[\tq\tqu]}(l_{\tq\tqu}) S_{\tqu}^T\overline{\g}^{0^-}_{[\tq\tqu]}(-k_{\tq\tq})S_{\tq}(l_{\tq})\D^{0^+}_{[\tq\tqu]}(l_{\tq\tqu})\}}\\
\nn & M_{\nu}^{62}=t^{\tq T}t^{\{\tq\tq\}_{1^+}}t^{[\tq\tqu]_{0^-}T}t^{\tqu}\\
\nn &\times\{\textcolor{magenta}{\g^{1^+}_{\{\tq\tq\},\rho}(l_{\tq\tq}) S_{\tq}^T\overline{\g}^{0^-}_{[\tq\tqu]}(-k_{\tq\tqu})S_{\tqu}(l_{\tqu})\D^{1^+}_{\{\tq\tq\},\rho\nu}(l_{\tq\tq})}\}\\
\nn &M_{\nu}^{63}=t^{\tq T}t^{\{\tq\tqu\}_{1^+}}t^{[\tq\tqu]_{0^-}T}t^{\tq}\\
\nn &\times\{\textcolor{magenta}{\g^{1^+}_{\{\tq\tqu\},\rho}(l_{\tq\tqu}) S_{\tqu}^T\overline{\g}^{0^-}_{\{\tq\tqu\}}(-k_{\tq\tqu})S_{\tq}(l_{\tq})\D^{1^+}_{\{\tq\tqu\},\rho\nu}(l_{\tq\tqu})}\}
\\
\nn & M^{66}_{\mu\nu}=t^{\tq T}t^{[\tq\tqu]_{0^-}}t^{[\tq\tqu]_{0^-}T}t^{\tq}\\
\nn&\times\{\textcolor{magenta}{\g^{0^-}_{[\tq\tqu]}(l_{\tq\tqu}) S_{\tq}^T\overline{\g}^{0^-}_{[\tq\tqu]}(-k_{\tq\tqu})S_{\tqu}(l_{\tqu})\D^{0^-}_{[\tq\tqu]}(l_{\tq\tqu})}\}\\
\nn & M_{\nu}^{67}=t^{\tq T}t^{\{\tq\tq\}_{1^-}}t^{[\tq\tqu]_{0^-}T}t^{\tqu}\\
\nn &\times\{\textcolor{magenta}{\g^{1^-}_{\{\tq\tq\},\rho}(l_{\tq\tq}) S_{\tq}^T\overline{\g}^{0^-}_{[\tq\tqu]}(-k_{\tq\tqu})S_{\tqu}(l_{\tqu})\D^{1^-}_{\{\tq\tq\},\rho\nu}(l_{\tq\tq})}\}\\
\nn & M_{\nu}^{68}=t^{\tq T}t^{\{\tq\tqu\}_{1^-}}t^{[\tq\tqu]_{0^-}T}t^{\tq}\\
\nn & \times\{\textcolor{magenta}{\g^{1^-}_{\{\tq\tqu\},\rho}(l_{\tq\tqu}) S_{\tqu}^T\overline{\g}^{0^-}_{[\tq\tqu]}(-k_{\tq\tqu})S_{\tq}(l_{\tq})\D^{1^-}_{\{\tq\tqu\},\rho\nu}(l_{\tq\tqu})}\}\\
\nn &M_{\mu}^{71}=t^{\tqu T}t^{[\tq\tqu]_{0^+}}t^{\{\tq\tq\}_{1^-}T}t^{\tq}\\
\nn &\times\{\textcolor{magenta}{\{\g^{0^+}_{[\tq\tqu]}(l_{\tq\tqu}) S_{\tq}^T\overline{\g}^{1^-}_{\{\tq\tq\},\mu}(-k_{\tq\tq})S_{\tq}(l_{\tq})\D^{0^+}_{[\tq\tqu]}(l_{\tq\tqu})\}}\\
\nn & M_{\mu\nu}^{72}=t^{\tqu T}t^{\{\tq\tq\}_{1^+}}t^{\{\tq\tq\}_{1^-}T}t^{\tqu}\\
&\nn \times\{\textcolor{magenta}{\g^{1^+}_{\{\tq\tq\},\rho}(l_{\tq\tq}) S_{\tq}^T\overline{\g}^{1^-}_{\{\tq\tq\},\mu}(-k_{\tq\tq})S_{\tq}(l_{\tq})\D^{1^+}_{\{\tq\tq\},\rho\nu}(l_{\tq\tq})}\}\\
\nn & M_{\mu\nu}^{73}=t^{\tqu T}t^{\{\tq\tqu\}_{1^+}}t^{\{\tq\tq\}_{1^-}T}t^{\tq}\\
\nn &\times\{\textcolor{magenta}{\g^{1^+}_{\{\tq\tqu\},\rho}(l_{\tq\tqu}) S_{\tq}^T\overline{\g}^{1^-}_{\{\tq\tq\},\mu}(-k_{\tq\tq})S_{\tq}(l_{\tq})\D^{1^+}_{\{\tq\tqu\},\rho\nu}(l_{\tq\tqu})}\}\\
\nn & M_{\mu}^{76}=t^{\tqu T}t^{[\tq\tqu]_{0^-}}t^{\{\tq\tq\}_{1^-}T}t^{\tq}\\
\nn&\times\{\textcolor{magenta}{\g^{0^-}_{[\tq\tqu]}(l_{\tq\tqu}) S_{\tq}^T\overline{\g}^{1^-}_{\{\tq\tq\},\mu}(-k_{\tq\tq})S_{\tq}(l_{\tq})\D^{0^-}_{[\tq\tqu]}(l_{\tq\tqu})}\}\\
\nn & M_{\mu\nu}^{77}=t^{\tqu T}t^{\{\tq\tq\}_{1^-}}t^{\{\tq\tq\}_{1^-}T}t^{\tqu}\\
&\nn \times\{\textcolor{magenta}{\g^{1^-}_{\{\tq\tq\},\rho}(l_{\tq\tq}) S_{\tq}^T\overline{\g}^{1^-}_{\{\tq\tq\},\mu}(-k_{\tq\tq})S_{\tq}(l_{\tq})\D^{1^-}_{\{\tq\tq\},\rho\nu}(l_{\tq\tq})}\}\\
\end{align*}
\begin{align*}
\nn &M_{\mu\nu}^{78} =t^{\tqu T}t^{\{\tq\tqu\}_{1^-}}t^{\{\tq\tq\}_{1^-}T}t^{\tq}
\\ \nn &\times\{\textcolor{magenta}{\g^{1^-}_{\{\tq\tqu\},\rho}(l_{\tq\tqu}) S_{\tq}^T\overline{\g}^{1^-}_{\{\tq\tq\},\mu}(-k_{\tq\tq})S_{\tq}(l_{\tq})\D^{1^-}_{\{\tq\tqu\},\rho\nu}(l_{\tq\tqu})}\}\\
\nn & M_{\mu}^{81}=t^{\tq T}t^{[\tq\tqu]_{0^+}}t^{\{\tq\tqu\}_{1^-}T}t^{\tq}\\
\nn &\times\{\textcolor{magenta}{\g^{0^+}_{[\tq\tqu]}(l_{\tq\tqu}) S_{\tqu}^T\overline{\g}^{1^-}_{\{\tq\tqu\},\mu}(-k_{\tq\tqu})S_{\tq}(l_{\tq})\D^{0^+}_{[\tq\tqu]}(l_{\tq\tqu})}\}\\
\nn& M_{\mu\nu}^{82}=t^{\tq T}t^{\{\tq\tq\}_{1^+}}t^{\{\tq\tqu\}_{1^-}T}t^{\tqu}\\
\nn &\times\{\textcolor{magenta}{\g^{1^+}_{\{\tq\tq\},\rho}(l_{\tq\tq}) S_{\tq}^T\overline{\g}^{1^-}_{\{\tq\tqu\},\mu}(-k_{\tq\tqu})S_{\tqu}(l_{\tqu})\D^{1^+}_{\{\tq\tq\},\rho\nu}(l_{\tq\tq})}\}\\
\nn & M_{\mu\nu}^{83}=t^{\tq T}t^{\{\tq\tqu\}_{1^+}}t^{\{\tq\tqu\}_{1^-}T}t^{\tq}
\\ \nn & \times\{\textcolor{magenta}{\g^{1^+}_{\{\tq\tqu\},\rho}(l_{\tq\tqu}) S_{\tq}^T\overline{\g}^{1^-}_{\{\tq\tqu\},\mu}(-k_{\tq\tqu})S_{\tq}(l_{\tq})\D^{1^+}_{\{\tq\tqu\},\rho\nu}(l_{\tq\tqu})}\}\\
\nn & M_{\mu}^{86}=t^{\tq T}t^{[\tq\tqu]_{0^-}}t^{\{\tq\tqu\}_{1^-}T}t^{\tq}\\
\nn &\times\{\textcolor{magenta}{\g^{0^-}_{[\tq\tqu]}(l_{\tq\tqu}) S_{\tq}^T\overline{\g}^{1^-}_{\{\tq\tqu\},\mu}(-k_{\tq\tqu})S_{\tqu}(l_{\tqu})\D^{0^-}_{[\tq\tqu]}(l_{\tq\tqu})}\\
\nn& M_{\mu\nu}^{87}=t^{\tq T}t^{\{\tq\tq\}_{1^-}}t^{\{\tq\tqu\}_{1^-}T}t^{\tqu}\\
\nn &\times\{\textcolor{magenta}{\g^{1^-}_{\{\tq\tq\},\rho}(l_{\tq\tq}) S_{\tq}^T\overline{\g}^{1^-}_{\{\tq\tqu\},\mu}(-k_{\tq\tqu})S_{\tqu}(l_{\tqu})\D^{1^-}_{\{\tq\tq\},\rho\nu}(l_{\tq\tq})}\}\\
\nn & M_{\mu\nu}^{88}=t^{\tq T}t^{\{\tq\tqu\}_{1^-}}t^{\{\tq\tqu\}_{1^-}T}t^{\tq}\\
\nn &\times\{\textcolor{magenta}{\g^{1^-}_{\{\tq\tqu\},\rho}(l_{\tq\tqu}) S_{\tq}^T\overline{\g}^{1^-}_{\{\tq\tqu\},\mu}(-k_{\tq\tqu})S_{\tq}(l_{\tq})\D^{1^-}_{\{\tq\tqu\},\rho\nu}(l_{\tq\tqu})}\}
\end{align*}
Note:
\bea \nn
\begin{array}{@{\extracolsep{0.1cm}}cccc}
M_\nu^{14}=M_\nu^{12}    & M_\nu^{15}=M_\nu^{13} &
M_\nu^{19}=M_\nu^{17}   & M_\nu^{110}=M_\nu^{18} \\
 \rule{0ex}{3.5ex}
M_{\mu\nu}^{24}=M_{\mu\nu}^{22}   & M_{\mu\nu}^{25}=M_{\mu\nu}^{23}  &
M_{\mu\nu}^{29}=M_{\mu\nu}^{27}   & M_{\mu\nu}^{210}=M_{\mu\nu}^{28} \\
 \rule{0ex}{3.5ex}
M_{\mu\nu}^{34}=M_{\mu\nu}^{32}   & M_{\mu\nu}^{35}=M_{\mu\nu}^{33}  &
M_{\mu\nu}^{39}=M_{\mu\nu}^{37}   & M_{\mu\nu}^{310}=M_{\mu\nu}^{38} \\
\end{array}
\eea
where $t^f$ are the flavour matrices and can be found in appendix \ref{app:Fla}. We observe that the following rows are equal: row $4=$ row $2$, row $5=$ row $3$, row $9= $row $7$ and row $10=$ row $8$.
\section{Flavor Diquarks}
\label{app:Fla}
We define the following set of flavor column matrices,
\begin{equation}\nn
t^{\tu}=\begin{pmatrix} 1  \\ 0  \\ 0 \\ 0 \\0 \end{pmatrix},\;\;\;\;
t^{\td}=\begin{pmatrix} 0  \\ 1  \\ 0  \\ 0\\0  \end{pmatrix},\;\;\;\;
t^{\ts}=\begin{pmatrix} 0  \\ 0  \\ 1 \\ 0 \\0  \end{pmatrix},\;\;\;\;
\end{equation}
\begin{equation}
t^{\tc}=\begin{pmatrix} 0  \\ 0  \\ 0 \\ 1\\0  \end{pmatrix},\;\;\;\;
t^{\tb}=\begin{pmatrix} 0  \\ 0  \\ 0 \\ 0\\1  \end{pmatrix},\;\;\;\;
\end{equation}
and
\bea \nn
t^f=\left[\begin{matrix} \nn
t^{\tq T}t^{\{\tqu \tq\}}t^{\{\tqu\tq\}T}t^{\tq} &&t^{\tq T}t^{\{\tq\tq\}}t^{\{\tqu\tq\}T}t^{\tqu},\\ \\
t^{\tqu T}t^{\{\tqu\tq\}}t^{\{\tq\tq\}T}t^{\tq} && t^{\tqu T}t^{\{\tq\tq\}}t^{\{\tq\tq\}T}t^{\tqu}
\end{matrix}\right] \,.
\eea
The flavor matrices for the diquarks are
\begin{equation}\nonumber
\begin{array}{cc}
\bf{t^{[\textcolor{red}{u}\textcolor{blue}{d}]}=\begin{pmatrix} 0 & 1 & 0 & 0 & 0 \\ -1 & 0 & 0  & 0& 0 \\ 0 &  0 & 0  & 0 & 0 \\ 0 &  0 & 0  & 0 & 0\\  0 & 0 & 0 & 0 & 0 \end{pmatrix}},
&
\bf{t^{[\textcolor{red}{u}\textcolor{green}{s}]}=\begin{pmatrix} 0 & 0 & 1& 0 & 0  \\ 0 & 0 & 0& 0& 0 \\ -1 &  0 & 0& 0 & 0\\ 0 &  0 & 0  & 0& 0\\  0 & 0 & 0 & 0 & 0 \end{pmatrix}},
\end{array}
\end{equation}
\begin{equation}\nonumber
\begin{array}{cc}
\bf{t^{[\textcolor{blue}{d}\textcolor{green}{s}]}=\begin{pmatrix} 0 & 0 & 0 & 0 & 0 \\ 0 & 0 & 1& 0 & 0  \\ 0 &  -1 & 0 & 0& 0   \\ 0 &  0 & 0  & 0& 0 \\  0 & 0 & 0 & 0 & 0 \end{pmatrix}},
&
\bf{t^{[\tu\tc]}=\begin{pmatrix} 0 & 0 & 0 & 1& 0  \\ 0 & 0 & 0& 0& 0   \\ 0 &  0 & 0 & 0& 0   \\ -1 &  0 & 0  & 0& 0\\  0 & 0 & 0 & 0 & 0 \end{pmatrix}},
\end{array}
\end{equation}
\begin{equation}\nonumber
\begin{array}{cc}
\bf{t^{\{\textcolor{red}{u}\textcolor{red}{u}\}}=\begin{pmatrix} \sqrt{2} & 0 & 0& 0 &0\\ 0 & 0 & 0& 0&0 \\ 0 &  0 & 0 & 0&0 \\ 0 &  0 & 0 & 0&0\\ 0 & 0 & 0& 0& 0  \end{pmatrix}},
&
\bf{t^{\{\textcolor{red}{u}\textcolor{blue}{d}\}}=\begin{pmatrix} 0 & 1 & 0 & 0&0\\ 1 & 0 & 0& 0&0 \\ 0 &  0 & 0& 0&0 \\ 0 &  0 & 0 & 0 &0\\ 0 & 0 & 0& 0& 0  \end{pmatrix}},
\end{array}
\end{equation}
\begin{equation}\nonumber
\begin{array}{cc}
\bf{t^{\{\tu\ts\}}=\begin{pmatrix} 0 & 0 & 1& 0&0  \\ 0 & 0 & 0& 0&0  \\ 1 &  0 & 0& 0 &0 \\ 0 &  0 & 0 & 0&0 \\ 0 & 0 & 0& 0& 0   \end{pmatrix}},
&
t^{\{\tu\tc\}}=\begin{pmatrix} 0 & 0 & 0& 1&0 \\ 0 & 0 & 0 & 0&0\\ 0 &  0 & 0 & 0&0\\  1 &  0 & 0 & 0&0\\ 0 & 0 & 0& 0& 0  \end{pmatrix},
\end{array}
\end{equation}
\begin{equation}\nonumber
\begin{array}{cc}
\bf{t^{\{\td\td\}}=\begin{pmatrix} 0 & 0 & 0 & 0 & 0 \\ 0 &  \sqrt{2} & 0& 0 & 0 \\ 0 &  0 & 0& 0& 0  \\ 0 & 0 & 0 & 0& 0\\ 0 & 0 & 0& 0& 0  \end{pmatrix}},
&
\bf{t^{\{\td\ts\}}}=\begin{pmatrix} 0 & 0 & 0& 0 & 0 \\ 0 & 0 & 1& 0 & 0\\ 0 &  1 & 0& 0 & 0\\ 0 & 0 & 0 & 0& 0\\ 0 & 0 & 0& 0& 0   \end{pmatrix},\;\;\;\;
\end{array}
\end{equation}
\begin{equation}\nonumber
\begin{array}{cc}
\bf{t^{\{\ts\ts\}}=\begin{pmatrix} 0 & 0 & 0& 0& 0 \\ 0 & 0 & 0& 0& 0 \\ 0 &  0 & \sqrt{2} & 0 & 0 \\ 0 & 0 & 0 & 0 & 0\\ 0 & 0 & 0& 0& 0  \end{pmatrix}},
&
\bf{t^{\{\tc\tc\}}}=\begin{pmatrix} 0 & 0 & 0& 0 & 0 \\ 0 & 0 & 0 & 0 & 0\\ 0 &  0 & 0 & 0 & 0\\  0 &  0 & 0 &  \sqrt{2} & 0\\ 0 & 0 & 0& 0& 0   \end{pmatrix},
\end{array}
\end{equation}
\begin{equation}\nonumber
\begin{array}{cc}
\bf{t^{[\td\tc]}}=\begin{pmatrix} 0 & 0 & 0& 0& 0 \\ 0 & 0 & 0 & 1& 0\\ 0 &  0 & 0 & 0& 0\\  0 &  -1& 0 & 0& 0 \\ 0 & 0 & 0& 0& 0    \end{pmatrix},
&
\bf{t^{[\ts\tc]}}=\begin{pmatrix} 0 & 0 & 0& 0 & 0 \\ 0 & 0 & 0 & 0& 0\\ 0 &  0 & 0 & 1& 0\\  0 &  0& -1 & 0& 0 \\ 0 & 0 & 0& 0& 0    \end{pmatrix},
\end{array}
\end{equation}
\begin{equation}\nonumber
\begin{array}{cc}
\bf{t^{\{\td\tc\}}}=\begin{pmatrix} 0 & 0 & 0& 0 & 0\\ 0 & 0 & 0 & 1& 0\\ 0 &  0 & 0 & 0& 0\\  0 &  1& 0 & 0& 0 \\ 0 & 0 & 0& 0& 0   \end{pmatrix},
&
\bf{t^{\{\ts\tc\}}}=\begin{pmatrix} 0 & 0 & 0& 0 & 0 \\ 0 & 0 & 0 & 0 & 0\\ 0 &  0 & 0 & 1 & 0\\  0 &  0& 1 & 0 & 0 \\ 0 & 0 & 0& 0& 0   \end{pmatrix}, \\[10ex]
\end{array}
\end{equation}
\begin{equation}\nonumber
\begin{array}{cc}
\bf{t^{[\tb\tu]}}=\begin{pmatrix} 0 & 0 & 0& 0& -1\\ 0 & 0 & 0 & 0& 0\\ 0 &  0 & 0 & 0& 0\\  0 &  0& 0 & 0& 0 \\ 1 & 0 & 0& 0& 0    \end{pmatrix},
&
\bf{t^{[\tb\td]}}=\begin{pmatrix} 0 & 0 & 0& 0 & 0 \\ 0 & 0 & 0 & 0& -1\\ 0 &  0 & 0 & 0& 0\\  0 &  0& 0 & 0& 0 \\ 0 & 1 & 0& 0& 0    \end{pmatrix},
\\
\end{array}
\end{equation}
\begin{equation}\nonumber
\begin{array}{cc}
\bf{t^{[\tb\tc]}}=\begin{pmatrix} 0 & 0 & 0& 0 & 0 \\ 0 & 0 & 0 & 0& 0\\ 0 &  0 & 0 & 0& 0\\  0 &  0& 0 & 0& -1 \\ 0 & 0 & 0& 1& 0    \end{pmatrix},
&
\bf{t^{[\tb\ts]}}=\begin{pmatrix} 0 & 0 & 0& 0 & 0 \\ 0 & 0 & 0 & 0& 0\\ 0 &  0 & 0 & 0& -1\\  0 &  0& 0 & 0& 0 \\ 0 & 0 & 1& 0& 0    \end{pmatrix},\;\;\;\;
\\
\end{array}
\end{equation}
\begin{equation}\nonumber
\begin{array}{cc}
\bf{t^{\{\tb\tb\}}}=\begin{pmatrix} 0 & 0 & 0& 0& 0\\ 0 & 0 & 0 & 0& 0\\ 0 &  0 & 0 & 0& 0\\  0 &  0& 0 & 0& 0 \\ 0 & 0 & 0& 0&  \sqrt{2}    \end{pmatrix}
&
\bf{t^{\{\tb\tu\}}}=\begin{pmatrix} 0 & 0 & 0& 0& 1\\ 0 & 0 & 0 & 0& 0\\ 0 &  0 & 0 & 0& 0\\  0 &  0& 0 & 0& 0 \\ 1 & 0 & 0& 0& 0 \end{pmatrix},\;\;\;\;
\\
\end{array}
\end{equation}
\begin{equation}\nonumber
\begin{array}{cc}
\bf{t^{\{\tb\td\}}}=\begin{pmatrix} 0 & 0 & 0& 0& 0\\ 0 & 0 & 0 & 0& 1\\ 0 &  0 & 0 & 0& 0\\  0 &  0& 0 & 0& 0 \\ 0 & 1 & 0& 0& 0   \end{pmatrix}
&
\bf{t^{\{\tb\tc\}}}=\begin{pmatrix} 0 & 0 & 0& 0& 0\\ 0 & 0 & 0 & 0& 0\\ 0 &  0 & 0 & 0& 0\\  0 &  0& 0 & 0& 1 \\ 0 & 0 & 0& 1& 0   \end{pmatrix},
\\
\end{array}
\end{equation}
\begin{equation}\nonumber
\begin{array}{cc}
\bf{t^{\{\tb\ts\}}}=\begin{pmatrix} 0 & 0 & 0& 0& 0\\ 0 & 0 & 0 & 0& 0\\ 0 &  0 & 0 & 0& 1\\  0 &  0& 0 & 0& 0 \\ 0 & 0 & 1& 0& 0
&
 \end{pmatrix}.
 \end{array}
\end{equation}
\vspace{0cm}

\bibliography{ccc-a}
\end{document}